\documentclass[usenatbib,useAMS]{mn2e}
\bibpunct{(}{)}{;}{a}{}{,}
\usepackage[dvips]{graphicx}
\usepackage{times}

\newcommand{\mic}{\hbox{$\mu$m}}

\newcommand{\hi}{\mbox{H\,{\sc i}}}
\newcommand{\tbgswarm}{\hbox{$T_\mathrm{W}^{\,\mathrm{BC}}$}}

\newcommand{\tbgscold}{\hbox{$T_\mathrm{C}^{\,\mathrm{ISM}}$}}
\newcommand{\fmu}{\hbox{$f_\mu$}}
\newcommand{\ldust}{\hbox{$L_{\mathrm{d}}^{\,\mathrm{tot}}$}}
\newcommand{\mdust}{\hbox{$M_{\mathrm{d}}$}}
\newcommand{\mgas}{\hbox{$M_{\mathrm{H}}$}}
\newcommand{\lsun}{\hbox{$L_{\sun}$}}
\newcommand{\msun}{\hbox{$M_{\sun}$}}

\newcommand{\xipahs}{\hbox{$\xi_\mathrm{PAH}^\mathrm{\,BC}$}}

\newcommand{\ximir}{\hbox{$\xi_\mathrm{MIR}^\mathrm{\,BC}$}}

\newcommand{\xiwarm}{\hbox{$\xi_\mathrm{W}^\mathrm{\,BC}$}}

\newcommand{\xicold}{\hbox{$\xi_\mathrm{C}^\mathrm{\,ISM}$}}

\newcommand{\mwarmbc}{\hbox{$M_\mathrm{W}^\mathrm{\,BC}$}}
\newcommand{\mwarmism}{\hbox{$M_\mathrm{W}^\mathrm{\,ISM}$}}
\newcommand{\mcoldism}{\hbox{$M_\mathrm{C}^\mathrm{\,ISM}$}}

\newcommand{\tauv}{\hbox{$\hat{\tau}_{V}$}}

\newcommand{\sfr}{\hbox{$\psi$}}
\newcommand{\ssfr}{\hbox{$\psi_{\mathrm S}$}}

\newcommand{\iras}{\hbox{\it IRAS}}
\newcommand{\galex}{\hbox{\it GALEX}}

\newcommand{\mstar}{\hbox{$M_{*}$}}

\title[Star formation activity and dust content in galaxies]
{New insight into the relation between star formation activity and
dust content in galaxies}

\author[E. da Cunha et al.]
{Elisabete~da~Cunha$^{1,2,3}$\thanks{E-mail: dacunha@physics.uoc.gr}, Celine~Eminian$^{4}$, St\'ephane~Charlot$^{2,3}$, J\'er\'emy~Blaizot$^{5}$
\\
$^{1}$Department of Physics, University of Crete, 71003 Heraklion, Greece; \\
IESL/Foundation for Research and Technology-Hellas, 71110 Heraklion, Greece. \\
$^{2}$UPMC Univ Paris 06, UMR7095, Institut d'Astrophysique de Paris, F-75014, Paris, France.\\
$^{3}$CNRS, UMR7095, Institut d'Astrophysique de Paris, F-75014, Paris, France.\\
$^{4}$Astronomy Centre, University of Sussex, Brighton BN1 9QH, UK.\\
$^{5}$Universit\'e de Lyon, Lyon, F-69003, France;\\
Universit\'e Lyon 1, Observatoire de Lyon, 9 avenue Charles
Andr\'e, Saint-Genis
Laval, F-69230, France;\\
CNRS, UMR 5574, Centre de Recherche Astrophysique de Lyon; Ecole
Normale Sup\'erieure de Lyon, Lyon, F-69007, France.}
\date{Accepted 2009 December 23.  Received 2009 December 13; in original form 2009 August 4}
\begin{document}

\maketitle

\begin{abstract}
We assemble a sample of 3258 low-redshift galaxies from the Sloan Digital
Sky Survey Data Release 6 (SDSS DR6) with complementary photometric 
observations by \galex, 2MASS and \iras\ at far-ultraviolet and infrared 
wavelengths. We use a recent, simple but physically motivated  model to 
interpret the observed spectral energy distributions of the galaxies in 
this sample in terms of statistical constraints on physical parameters
describing the star formation history and dust content. The focus on 
a subsample of 1658 galaxies with highest signal-to-noise ratio (S/N) 
observations enables us to investigate most clearly several strong correlations 
between various derived physical properties of galaxies. In particular, 
we find that the typical dust mass $\mdust$ of a galaxy forming stars
at a rate $\sfr$ can be estimated remarkably well using the formula 
$\mdust = (1.28\pm 0.02)\times10^7 \, (\sfr / \msun\ \mathrm{yr}^{-1})^{1.11
\pm 0.01} \,\msun$ over at least three orders of magnitude in both quantities. We also
find that the dust-to-stellar mass ratio, the ratio of dust mass to star
formation rate and the fraction of dust luminosity contributed by the 
diffuse interstellar medium (ISM) all correlate strongly with specific star
formation rate. A comparison with recent models of chemical and dust 
evolution of galaxies suggests that these correlations could arise, at 
least in part, from an evolutionary sequence.
As galaxies form stars, their ISM becomes enriched in dust, while the 
drop in gas supply makes the specific star formation rate decrease.
Interestingly, as a result, a young, actively star-forming galaxy with 
low dust-to-gas ratio may still be highly dusty
(in the sense of a high dust-to-stellar mass ratio)
because it contains large amounts of interstellar gas.
This may be important for the interpretation of the infrared emission from
young, gas-rich star-forming galaxies at high
redshift. The results presented in this paper should be especially useful
to improve the treatment of the ISM properties of galaxies in semi-analytic
models of galaxy formation.
Our study also provides a useful local reference for future statistical 
studies of the star formation and dust properties of galaxies at high redshifts.
\end{abstract}

\begin{keywords}
dust, extinction -- galaxies: ISM -- galaxies: stellar content -- galaxies: statistics -- galaxies: evolution.
\end{keywords}

\section{Introduction} \label{sect:intro}

Interstellar dust and star formation activity are strongly linked in
galaxies. Dust grains condense in the cold envelopes of evolved 
stars and in supernova ejecta. In return, they favour the formation
of molecular hydrogen, shield the newly-formed molecules from 
ultraviolet radiation and participate in the formation and cooling of 
molecular clouds, which collapse to form new stars.

Studies of the ultraviolet, optical and near-infrared emission
from large samples of local galaxies have shed some light on
the relation between star formation activity and dust content.
Such studies show that, in general, galaxies with the highest
star formation rates also suffer the highest ultraviolet 
and optical attenuation (e.g., \citealt{Kauffmann2003b,
Brinchmann2004}). However, observations at ultraviolet, optical
and near-infrared wavelengths set only limited constraints on
the dust content of galaxies. New insight into the dust 
properties requires observations at mid- and far-infrared
wavelengths, where the dust re-radiates the energy absorbed
from starlight. The first studies of this kind combined 
ultraviolet and optical observations of large samples of 
nearby galaxies with empirical estimates of the total 
infrared luminosity computed from observations with the
{\it Infrared Astronomical Satellite} (\iras) at 12, 25,
60 and 100~\mic\ (e.g.~\citealt{Wang1996,Hopkins2001,Sullivan2001,
Kong2004}). Other recent studies have taken this (largely 
empirical) approach a step further by developing consistent models
of galactic spectral energy distributions at ultraviolet, optical
and infrared wavelengths (e.g.~\citealt{Silva1998,Dopita2005}).
These sophisticated models combine the spectral evolution of stellar
populations with detailed calculations of the transfer of starlight
through the interstellar medium in galaxies. Such models, however,
are not optimised to interpret the observations of large samples of
galaxies.

Recently, \citet{daCunha2008} proposed a simple, physically motivated 
model to interpret consistently the ultraviolet, optical and infrared 
observations of large samples of galaxies in terms of statistical 
constraints on physical parameters pertaining to the stars, gas and 
dust. By design, this model is less sophisticated than other existing
radiative transfer models. In return, it allows one to interpret 
efficiently the spectral energy distributions of large samples of 
galaxies in terms of statistical constraints on a minimal set of 
adjustable physical parameters. \citet{daCunha2008} used this model
to derive median-likelihood estimates of the stellar mass, star 
formation rate, dust attenuation and dust mass from the observed 
ultraviolet, optical and infrared spectral energy distributions of
66 galaxies in the {\it Spitzer} Infrared Nearby Galaxy Sample (SINGS,
\citealt{Kennicutt2003}). For these galaxies, the dust-to-stellar mass
ratio appears to correlate strongly with specific star formation rate
(i.e.~the star formation rate divided by the total stellar mass), 
confirming the expectation that dust mass and star formation rate are
tightly related in galaxies.

In the present paper, we investigate further the relation between star
formation activity and dust content by studying the properties of a much
larger sample of 3258 star-forming galaxies, for which photometric
observations are available at ultraviolet, optical and infrared
wavelengths. We use the model of \citet{daCunha2008} to derive
median-likelihood estimates of several physical parameters of each
galaxy in this sample, such as the total stellar mass, dust mass
star formation rate and fractional contribution of different dust 
components to the total infrared luminosity. We then focus
on a sub-sample of 1658 galaxies with highest signal-to-noise
ratio (S/N) photometry from the far-ultraviolet to the far-infrared. 
The large size of this sample allows us to study in unprecedented 
detail the relations between specific star formation rate, dust mass
and dust-to-gas ratio already identified in the SINGS sample by 
\citet{daCunha2008}. We find that, in particular, the typical dust
mass $\mdust$ of a galaxy forming stars at a rate $\sfr$ can be 
remarkably well estimated using the formula $\mdust = (1.28\pm 
0.02)\times10^7 \, (\sfr / \msun\ \mathrm{yr}^{-1})^{1.11 \pm 0.01}
\,\msun$.

We interpret our results in the framework of recent models of
chemical and dust evolution of galaxies by \cite{Calura2008}. 
These models provide a means of predicting the time evolution
of the stellar and dust content of galaxies with different star
formation histories. We conclude from this that the relations
between specific star formation rate and dust content 
exhibited by the galaxies in our sample could be regarded, at
least in part, as an evolutionary sequence.

The paper is organised as follows. We first describe the matched
\galex-SDSS-2MASS-\iras\ sample in Section~\ref{dust:sample}. In
Section~\ref{dust:model}, we outline the method used to derived
statistical estimates of the total stellar mass, dust mass, star
formation rate and fractional contribution of different dust
components to the total infrared luminosity for each galaxy in 
this sample. We present the relations between these various 
derived quantities in Section~\ref{dust:rel_sfr_dust}. In 
Section~\ref{dust:discussion}, we discuss some potential biases 
of our approach. We also compare our results with the predictions
of different models to illustrate the implications of our study for
the evolution of the dust content of star-forming galaxies.
Section~\ref{dust:conclusion} summarises our conclusions.

%*****************************************************************************

\section{The galaxy sample} \label{dust:sample}

In this work we analyse a galaxy sample obtained by cross-correlating
the Sloan Digital Sky Survey Data Release 6 (SDSS DR6)
 main spectroscopic sample with photometric
catalogues at ultraviolet (from the {\it Galaxy Evolution Explorer}, \galex),
near-infrared (from the Two Micron All Sky Survey, 2MASS) and far-infrared
(from the {\it Infrared Astronomical Satellite}, \iras) wavelengths.

\subsection{Optical photometry}
\label{optic_phot}
The SDSS DR6 spectroscopic sample \citep{Adelman_McCarthy2008}
contains 792,680 galaxies with known redshifts to a Petrosian
$r$-band magnitude limit $r<17.77$. 
Optical photometry in the $ugriz$ bands (at 3557, 4825, 6261,
7672 and 9097~\AA, respectively) is available for these galaxies.
We use the SDSS `model magnitudes' in these bands, which reflect
the integrated light from the whole galaxy and are the best suited
to comparisons with total photometry from \galex, 2MASS and \iras\ 
at ultraviolet and infrared wavelengths. We stress that 
such consistent multi-wavelength photometry is a key
requirement in the interpretation of the spectral energy
distributions of the galaxies in Section~\ref{dust:model}.

Many fundamental parameters of the
galaxies of the SDSS DR6 spectroscopic sample
have been derived from the spectroscopic data (e.g.~
\citealt{Brinchmann2004}), and are publicly available in an online
database\footnote{http://pc-66.astro.up.pt/$\sim$jarle/GASS/}. We use
this information to restrict our sample to galaxies classified as
`star-forming' according to their emission lines. Active Galactic
Nuclei (AGNs) in the sample are eliminated when at least one of two
criteria are met: (i) the presence of broad emission lines from 
high-velocity gas around the AGN (e.g.~\citealt{Brinchmann2004});
(ii) the presence of narrow emission lines with intensity ratios
characteristic of an AGN in the Baldwin, Phillips \& Terlevich 
diagram (BPT; \citealt{Baldwin1981}), according to the criterion of
\citet{Kauffmann2003a}\footnote{The BPT diagram allows one to
distinguish star-forming galaxies from AGN on the basis of the
degree of ionisation of the gas by comparing the narrow-line
ratios [O~{\sc III}]$\lambda~5007$~\AA/H$\beta$ against [N~{\sc
II}]$\lambda~6584~$\AA/H$\alpha$.}. If the emission lines are
highly attenuated, the BPT diagnostic may fail to identify an 
obscured AGN (in Section \ref{dust:AGN}, we discuss the possible
contamination of our sample by optically thick AGN). Observations
in the X-rays or in the mid-infrared can be used to detect AGN
in this case.

\subsection{Ultraviolet photometry}
\label{uv_phot}
To supplement SDSS DR6 spectroscopic data with ultraviolet
measurements, we cross-correlate the sample with the
latest data release from \galex\ \citep{Martin2005, Morrissey2005}.
 \galex\ is an all-sky survey providing images of the
galaxies in two photometric bands: the far-ultraviolet ({\it
FUV}, at 1520~\AA) and the near-ultraviolet ({\it NUV}, at 2310~\AA).
To match the SDSS and \galex\ catalogues, we use a search radius of 
4\,arcsec around the SDSS position (data kindly provided in advance of
publication by David Schiminovich). We require a detection in at least
one of the two {\it FUV} and {\it NUV} bands.  We retain galaxies which
have a single \galex\ detection within the search radius, i.e., we
eliminate objects for which two or more \galex\ detections may exist
for the same SDSS source.
We adopt \galex\ FUV and NUV `automags', which use a Kron aperture defined by
the profile of each galaxy in each band, and are designed to recover
the total flux.

At ultraviolet wavelengths, Galactic foreground
extinction is particularly important. Therefore, we apply
Galactic reddening corrections to the \galex\
magnitudes \citep{Seibert2005}: $A_\mathrm{FUV}=8.29~E(B-V)$ and
$A_\mathrm{NUV}=8.18~E(B-V)$, where the colour excess $E(B-V)$ is
determined from the dust reddening maps of \citet{Schlegel1998}.
Although observations at longer wavelengths are less affected by Galactic extinction, we also correct the
observed optical magnitudes of Section~\ref{optic_phot} and near-infrared magnitudes
of Section~\ref{nir_phot} using the same method.

\subsection{Near-infrared photometry}
\label{nir_phot}
We further supplement our sample with near-infrared photometry in
the $JHK_s$ bands (at 1.25, 1.65 and 2.17~\mic, respectively)
from the 2MASS \citep{Skrutskie2006} All Sky
Extended Source Catalog (XSC). We cross-identify the SDSS DR6
spectroscopic sample and the 2MASS XSC within a search radius of 5
arcsecs around the SDSS coordinates. Only sources with no artefacts
and which are not in close proximity to a large nearby galaxy are retained.
Concerning the photometry, we follow the recommendations in
the User's Guide to the 2MASS All-Sky Data
Release\footnote{http://www.ipac.caltech.edu/2mass/releases/allsky/doc/} to include 
most of the integrated flux from the galaxies while providing accurate colours: 
we adopt K20 fiducial isophotal elliptical aperture magnitudes for
galaxies with $K_s < 13$, and fixed 7~arcsec circular
aperture magnitudes for fainter galaxies.

\subsection{Mid- and far-infrared photometry}
\label{fir_phot}
We now turn to the most original feature of our sample: the
inclusion of mid- and far-infrared photometry obtained with
\iras\ \citep{Beichman1988}. Observations with
\iras\ are rather limited in sensitivity, but they have the
advantage of providing an all-sky survey at 12, 25, 60 and
100~\mic.\footnote{In practice, most measurements at 12 and 25~\mic\ are
upper limits on the flux density, and we do not include them in our
spectral fits. Only about 3 per cent of the galaxies in the sample 
are detected in all 4 \iras\ bands.} We cross-correlate our
sample with both the \iras\ Point Source Catalogue
(PSCz; \citealt{Saunders2000}) and the \iras\ Faint Source
Catalogue v2.0 (FSC; \citealt{Moshir1989}). The PSCz catalogue is
a complete and uniform galaxy catalogue assembled from the \iras\
Point Source Catalogue \citep{Beichman1988} and supplemented by
various redshift surveys. It contains 15,411 galaxies with
measured redshifts to a depth of 0.6~Jy at 60~\mic. The Faint
Source Catalogue contains 173,044 sources to a depth of roughly
0.25~Jy at 60~\mic. Low Galactic latitude regions ($|b| < 20^{\circ}$) are
excluded from this catalogue because of the contamination by
foreground Galactic sources at this detection limit. For both
catalogues, we rely on the quoted data quality flags to
select sources having a good flux quality at 60~\mic\
and at least a moderate flux quality at 100~\mic;
detections at 12 and 25~\mic\ are not strictly required. To
increase the reliability of the sample, we exclude observations
which are confusion-limited or contaminated by cirrus emission.

We cross-correlate the SDSS sample with the \iras\ PSCz
and \iras\ FSC catalogues using the same criteria as 
\citet{Pasquali2005}, i.e., we discard \iras\ sources with
more than one SDSS-matched object within a search radius of 
30 arcsec. For the PSCz catalogue (with redshift information), 
this criterion appears to cause only about 1~per cent of the
infrared sources to be assigned the wrong SDSS counterpart.
The contamination may be higher in the case of the lower-quality
FSC catalogue. Using simulated data, \citet{Pasquali2005} estimate
that the percentage of wrong cross-identifications is at most
1.5~per cent for a matching radius of 30 arcsec.

\subsection{Final sample}

\begin{figure}
\begin{center}
\includegraphics[width=0.45\textwidth,angle=90]{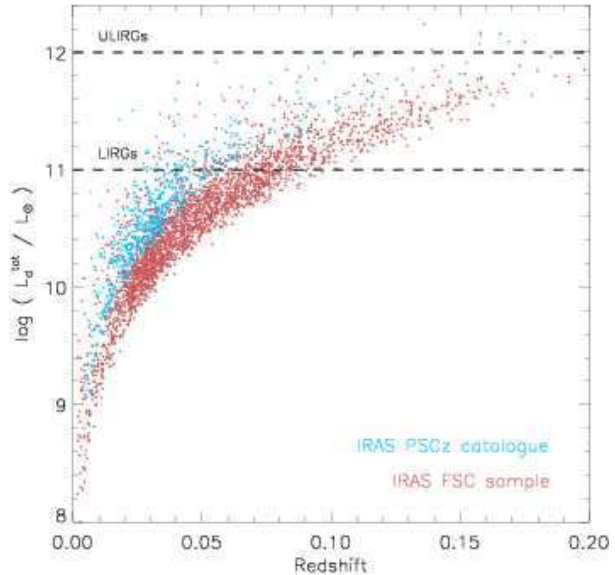}
\caption{Total infrared luminosity \ldust\ from 
equation~(\ref{Helou1988_1}) plotted against redshift for 
galaxies in the matched \galex-SDSS-2MASS-\iras\ described in
Section~\ref{dust:sample}. Galaxies belonging to the \iras\ PSCz
catalogue (blue symbols) are distinguished from those belonging to
the \iras\ FSC catalogue (red symbols).}
\label{fig1}
\end{center}
\end{figure}

\begin{figure}
\begin{center}
\includegraphics[width=0.45\textwidth,angle=90]{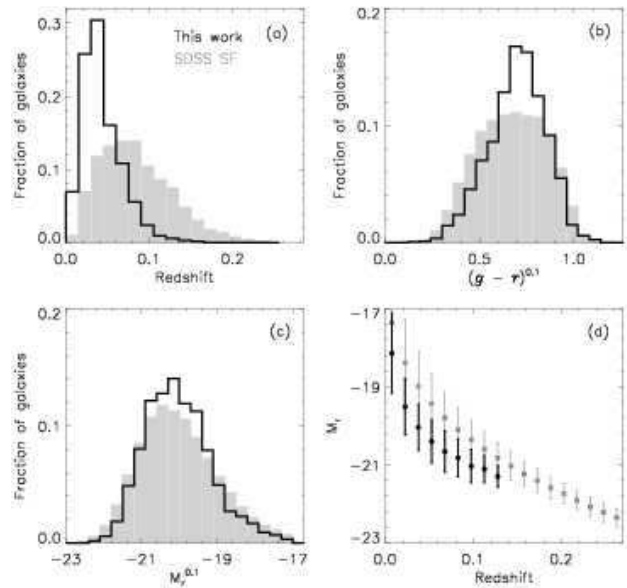}
\caption{Properties of the SDSS DR6 star-forming galaxies (in
grey) and the matched \galex-SDSS-2MASS-\iras\ subsample
considered in this work (in black). The histograms show the normalised
distributions of the following parameters: (a) Redshift, $z$; (b)
Galaxy $g-r$ colour, $k$-corrected to redshift $z = 0.1$; (c)
Absolute $r$-band model magnitude,
$k$-corrected to redshift $z = 0.1$,
M$_r^{\,\tiny{0.1}}$. Panel (d) shows
the mean $r$-band absolute magnitude, M$_r$, in different redshift bins for both samples
(the error bars represent the standard deviation $\pm \sigma$
in each bin).} \label{fig:dust2}
\end{center}
\end{figure}

Our final \galex-SDSS-2MASS-\iras\ sample is composed of 3258 galaxies
from the PSCz and FSC catalogues at redshifts $z\leq 0.22$
(see Fig.~\ref{fig:dust2}a). To summarise, the galaxies in this sample have:
\begin{itemize}
\item ultraviolet fluxes in at least one of the {\it FUV} and {\it NUV} \galex\ bands;
\item optical $ugriz$ fluxes from SDSS;
\item near-infrared $JHK_s$ fluxes from 2MASS;
\item 60- and 100-\mic\ flux densities from \iras\ (some galaxies 
also have 12- and  25-\mic\ measurements).
\end{itemize}
We note that, for consistency, all fluxes used in our study are designed to be {\it 
total fluxes}, which do not require aperture corrections.
For this reason, we also choose not to include SDSS spectroscopic
information available for these galaxies, which is restricted
to the inner 3-arcsec diameter aperture sampled by the SDSS
spectroscopic fibre aperture. At the low redshifts of our galaxies,
this limited spatial sampling could severely bias estimates of the
star formation rate and dust content \citep{Kewley2005}. In 
addition, spectroscopic quantities pertaining to a restricted 
central area cannot be compared directly with multi-wavelength 
photometric quantities describing the whole galaxy.

It is instructive to examine the typical infrared luminosity \ldust\
of the galaxies in our sample. For this purpose, we compute \ldust\ 
from the 60- and 100-\mic\ \iras\ flux densities $F_\nu^{60}$ and
$F_\nu^{100}$ using the empirical formula \citep{Helou1988}
\begin{equation}
L_\mathrm{d}^\mathrm{\,tot} = F_\mathrm{c}~L_\mathrm{FIR}\,,
\label{Helou1988_1}
\end{equation}
where
\begin{equation}
L_\mathrm{FIR}=1.26\times10^{-14}
(2.58\,F_{\nu}^{60}+F_{\nu}^{100})\,4\pi d_\mathrm{L}^{2}\,.
\label{Helou1988_2}
\end{equation}
Here, $d_\mathrm{L}$ is the luminosity distance in m, $F_\nu^{60}$
and $F_\nu^{100}$ are in Jy, and \ldust\ and $L_\mathrm{FIR}$ are
in W. We compute the correction factor $F_c$ 
to obtain \ldust\ from
$L_\mathrm{FIR}$ in equation~(\ref{Helou1988_1}) following the empirical
prescription of \cite{Helou1988}, which depends on the 
$F_{\nu}^{60}/F_{\nu}^{100}$ ratio, and assuming a dust emissivity index
$\beta=2$. The resulting median correction for our sample is $F_c=1.35$.

In Fig.~\ref{fig1}, we plot the total dust infrared luminosity 
\ldust\ computed in this way as a function of redshift. A
large fraction of the galaxies
in our sample (about 22 per cent) have $\ldust > 10^{11}$~\lsun.
Such galaxies are usually referred to as `luminous infrared
galaxies' (LIRGs). About 1 per cent of the galaxies of our sample have $\ldust >
10^{12}$~\lsun. These galaxies with extremely high infrared luminosities
are usually referred to as `ultra-luminous infrared galaxies'
(ULIRGs). This type of galaxies has been the object of extensive studies (e.g.~
\citealt{Soifer1987,Veilleux1995,Veilleux1999,Rigopoulou1999,Cao2006,Armus2007}).
In the local universe, most ULIRGs are observed to be the results
of interactions and mergers. Visual inspection of the SDSS optical images
confirms that ULIRGS in our sample also have disturbed morphologies.

In Fig.~\ref{fig:dust2}, we compare the properties of our
sample with the overall properties of the SDSS star-forming galaxy
sample. Fig.~\ref{fig:dust2}a shows that galaxies in our sample tend to
lie at lower redshifts than the bulk of SDSS star-forming galaxies.
This results from the required detection by low-sensitivity \iras\
observations. Fig.~\ref{fig:dust2}b further shows that galaxies in our
sample tend to have typically redder $g-r$ colours than SDSS 
star-forming galaxies. This is most probably a consequence of the
higher dust content of our galaxies, which makes them detectable
by \iras. In fact, \citet[][ see their figure
19]{Obric2006} point out that SDSS galaxies detected by \iras\
have systematically higher dust attenuation than the average SDSS
star-forming galaxy. The difference in the distributions of $r$-band
absolute magnitude M$_r$ in Fig.~\ref{fig:dust2}c is more subtle to
interpret. This is illustrated by Fig.~\ref{fig:dust2}d, where we plot
M$_{r}$ in different redshift bins for both samples. At redshifts $z<0.12$,
galaxies in our sample are typically brighter than SDSS star-forming 
galaxies. This is because SDSS galaxies with intrinsically faint $r$-band
magnitudes are too faint in the infrared to be detected by \iras. At 
redshifts $z>0.12$, the SDSS and \iras\ detection limits both correspond
to similarly bright galaxies.

\section{Statistical constraints on physical parameters}
\label{dust:model}

In this Section, we use the simple model of \citet{daCunha2008}
to extract star formation histories and dust contents from the
ultraviolet, optical and infrared observations of the galaxy 
sample described in Section~\ref{dust:sample}. We first briefly
summarise the model. Then, we describe the statistical approach
used to derive median-likelihood estimates of physical parameters
from the available data.

\subsection{Description of the model}
\label{dust:model_des}

The simple, physically motivated model of \citet{daCunha2008}
allows us to interpret the mid- and far-infrared spectral energy
distributions of galaxies consistently with the emission at
ultraviolet, optical and near-infrared wavelengths. We briefly
recall the main features of this model.

We compute the emission by stars in galaxies using the latest
version of the \citet{Bruzual2003} population synthesis code
(Charlot \& Bruzual, in preparation). This code predicts the
spectral evolution of stellar populations in galaxies from
far-ultraviolet to far-infrared wavelengths and at ages between
$1\times10^5$ and $2\times10^{10}$~yr, for different
metallicities, initial mass functions (IMFs) and star formation
histories. In this work, we adopt the \citet{Chabrier2003}
Galactic-disc IMF.

The emission from stars is attenuated using the simple
two-component dust model of \citet{Charlot2000}. This accounts for
the fact that stars are born in dense molecular clouds with
typical lifetimes of $10^7$~yr; at later ages, stars migrate to
the ambient (diffuse) ISM. Thus, the light produced by stars
younger than $10^7$~yr is attenuated by dust in the birth clouds
and in the ambient ISM, while the light produced by older stars
is attenuated only by dust in the ambient ISM. The model of 
\citet{Charlot2000} uses an `effective absorption' curve for each
component, $\hat \tau_\lambda \propto \lambda^{-n}$. The slope $n$
reflects both the optical properties and the spatial distribution
of the dust. Following \citet{Charlot2000}, we adopt for the ambient
ISM
\begin{equation}
\hat\tau_\lambda^\mathrm{\,ISM}=\mu \hat\tau_V (\lambda/5500~\mathrm{\AA})^{-0.7}\,,
\label{tau_ism}
\end{equation}
where $\hat \tau_V$ is the total effective $V$-band absorption
optical depth of the dust seen by young stars inside birth clouds,
and $\mu = \hat\tau_V^\mathrm{\,ISM}/(\hat\tau_V^\mathrm{\,BC} +
\hat\tau_V^\mathrm{\,ISM})$ is the fraction of this contributed by
dust in the ambient ISM. For the stellar birth clouds, we adopt:
\begin{equation}
\hat\tau_\lambda^\mathrm{\,BC} = (1-\mu) \hat\tau_V (\lambda/5500~\mathrm{\AA})^{-1.3}\,.
\label{tau_bc}
\end{equation}
We use this prescription to compute the total energy absorbed by
dust in the birth clouds and in the ambient ISM; this energy is
re-radiated by dust at infrared wavelengths. By analogy with
\citet{daCunha2008}, we define the total dust luminosity
re-radiated by dust in the birth clouds and in the ambient ISM as
$L_\mathrm{d}^\mathrm{\,BC}$ and $L_\mathrm{d}^\mathrm{\,ISM}$,
respectively. The total luminosity emitted by dust in the galaxy
is then
\begin{equation}
L_\mathrm{d}^\mathrm{\,tot} = L_\mathrm{d}^\mathrm{\,BC} + L_\mathrm{d}^\mathrm{\,ISM} \,.
\label{ldust}
\end{equation}

We distribute $L_\mathrm{d}^\mathrm{\,BC}$ and $L_\mathrm{d}^\mathrm{\,ISM}$
in wavelength over the range from 3 to 1000~\mic\ using 4 main components
(see \citealt{daCunha2008} for detail):
\begin{itemize}
\item the emission from polycyclic aromatic hydrocarbons (PAHs;
i.e.~mid-infrared emission features), \item the mid-infrared
continuum emission from hot dust with temperatures in the range
130--250~K, \item the emission from warm dust in thermal
equilibrium with adjustable temperature in the range 30--60~K,
\item the emission from cold dust in thermal equilibrium with
adjustable temperature in the range 15--25~K.
\end{itemize}
In stellar birth clouds, the relative contributions to
$L_\mathrm{d}^\mathrm{\,BC}$ by PAHs, the hot mid-infrared 
continuum and warm dust are kept as adjustable parameters.
These clouds are assumed not to contain any cold dust. In the
ambient ISM, the contribution to $L_\mathrm{d}^\mathrm{\,ISM}$
by cold dust is kept as an adjustable parameter. The relative
proportions of the other 3 components are fixed to the 
values reproducing the mid-infrared cirrus emission of the Milky
Way. \citet{daCunha2008} find that the above minimum number of 
components is required to account for the infrared spectral energy
distributions of galaxies in a wide range of star formation
histories.

\subsection{Median-likelihood estimates of physical parameters}
\label{dust:method}

The model summarised in Section~\ref{dust:model_des} allows us to 
derive statistical estimates of galaxy physical parameters, such as
the stellar mass, star formation rate and dust mass, from simultaneous
fits of ultraviolet, optical and infrared observations. To achieve 
this, we adopt a Bayesian approach similar to that used by 
\cite{daCunha2008} to interpret the spectral energy distributions
of SINGS galaxies.

\subsubsection{Model library}
\label{library}

We build large libraries of stochastic models at different redshifts
$z=0.00$, 0.05, 0.10, 0.15 and 0.20.

At each redshift, we generate a random library of stellar
population models for wide ranges of star formation histories,
metallicities and dust contents. Each star formation history is
parameterised in terms of an underlying continuous model with
exponentially declining star formation rate, on top of which are
superimposed random bursts (see also \citealt{Kauffmann2003b}).
The models are distributed uniformly in metallicity between 0.2
and 2 times solar. The attenuation by dust is randomly sampled by
drawing the total effective $V$-band absorption optical depth,
\tauv, between 0 and 6, and the fraction of this contributed by
dust in the ambient ISM, $\mu$, between 0 and 1 (see
\citealt{daCunha2008} for more details on the prior distributions
of these parameters). For each model, we compute the fraction of
the total energy absorbed by dust in the
diffuse ISM, $f_\mu$. We also compute the specific star formation
rate averaged over the last $t_8=10^8$~yr,
\begin{equation}
\ssfr(t)=\frac{\int_{t-t_8}^t dt' \sfr(t')}{t_8 M_\ast(t)}\,,
\label{eq:ssfr}
\end{equation}
where $M_\ast(t)$ is the stellar mass at time $t$.

In parallel, at each redshift, we generate a random library of
infrared spectra for wide ranges of dust temperatures and
fractional contributions by different dust components to the total
infrared luminosity. The fraction of total dust luminosity 
contributed by the diffuse ISM, $f_\mu =
L_\mathrm{d}^\mathrm{\,ISM}/ L_\mathrm{d}^\mathrm{\,tot}$, and the
fractional contribution by warm dust in thermal equilibrium to the
total dust luminosity of the birth clouds, \xiwarm, are uniformly
distributed between 0 and 1; the fractions of luminosity
contributed by the other components in the birth clouds, \ximir\
(hot mid-infrared continuum) and \xipahs\ (PAHs), are randomly
drawn from prior distributions such that $\xiwarm+\ximir+\xipahs = 1$.
We distribute uniformly the temperature of warm
dust in thermal equilibrium in the birth clouds, \tbgswarm,
between 30 and 60~K, and the temperature of cold dust in thermal
equilibrium in the diffuse ISM, \tbgscold, between 15 and 25~K.
Finally, we distribute uniformly the contribution by cold dust to
the total infrared luminosity of the ISM, \xicold, between 0.5 and
1.

We compute the dust mass associated to each model in the library
of infrared spectra as
\begin{equation}
\mdust=1.1\,(\mwarmbc + \mwarmism + \mcoldism)\,,
\label{eq:mdust}
\end{equation}
where \mwarmbc, \mwarmism\ and \mcoldism\ are the masses of the
dust components in thermal equilibrium (warm dust in the birth
clouds and the ambient ISM and cold dust in the ambient ISM).
The multiplying factor 1.1 accounts for the small contribution
by stochastically heated dust (see \citealt{daCunha2008} for detail).
The mass $M_\mathrm{d}(T_\mathrm{d})$ of dust in thermal equilibrium
at the temperature $T_\mathrm{d}$ is estimated from the corresponding
far-infrared luminosity, $L_\lambda^{\,T_\mathrm{d}}$, using the 
relation \citep{Hildebrand1983}
\begin{equation}
L_\lambda^{\,T_\mathrm{d}}=4\pi\, M_\mathrm{d}(T_\mathrm{d})\,\kappa_\lambda\,B_\lambda(T_\mathrm{d})\,,
\label{eq:h83}
\end{equation}
where $\kappa_\lambda$ is the dust mass absorption coefficient and
$B_\lambda(T_\mathrm{d})$ the Planck function of temperature
$T_\mathrm{d}$. Following \cite{daCunha2008}, we adopt
$\kappa_\lambda \propto \lambda^{-\beta}$, with $\beta=1.5$ for warm dust
and $\beta=2.0$ for cold dust, normalised to $\kappa_{850}=
0.77$~g$^{-1}$~cm$^2$ at 850~\mic\ \citep{Dunne2000}.

We combine the stochastic libraries of attenuated stellar spectra
and dust emission spectra at each redshift by associating models 
corresponding to similar values of \fmu\ (to within some uncertainty interval
$\delta f_\mu=0.15$) in the two libraries, which we scale to the same
total dust luminosity \ldust. For each combined spectrum, we compute
the synthetic photometry in the \galex\ $FUV$ and $NUV$, SDSS $ugriz$, 
2MASS $JHK_s$ and \iras\ 12-, 25-, 60- and 100-\mic\ bands.

\subsubsection{Corrections applied to the observed fluxes}
\label{corrections}
In the redshift range of the sample described in 
Section~\ref{dust:sample}, prominent optical nebular emission lines
can significantly affect the observed galaxy fluxes in the
SDSS $gri$ bands. The model spectra we use to interpret these data
do not include nebular emission lines. Therefore, to interpret the
optical fluxes of observed galaxies with these models, we first
need to correct the observed SDSS $gri$ magnitudes for
potential contamination by nebular emission lines (e.g.,
\citealt{Kauffmann2003b}). We use the corrections inferred by
Jarle Brinchmann (private communication) from fits of the 
stellar continuum emission of each SDSS optical spectrum with
\cite{Bruzual2003} models (we assume for simplicity that the 
correction derived in this way within the aperture sampled by
the SDSS fibre applies to the galaxy as a whole).

We also compute {\it k}-corrections to the ultraviolet, optical
and near-infrared magnitudes of each galaxy in the sample. To
minimise these, we {\it k}-correct the magnitudes from the
galaxy redshift to the closest redshift of the model grid 
described in Section~\ref{library}, i.e., $z=0.00$, 0.05, 0.10,
0.15 or 0.20 (we use the version v3 of the {\sc kcorrect} code
of \citealt{Blanton2003}). This procedure cannot be extended to 
the \iras\ 12-, 25-, 60- and 100-\mic\ flux densities, to which
we do not apply any {\it k}-correction. We do not expect this to have
any noticeable influence on our results, given the large effective
width of the \iras\ filter response functions and the relatively
large observational uncertainties in these bands.

To account for the uncertainties linked to the $k$-correction and
emission-line correction, we add the following errors to the quoted
flux uncertainties: 2 per cent for \galex, 2MASS, and SDSS $z$
bands, and 1.5 per cent for the SDSS $gri$ bands. For the less
accurate SDSS $u$-band photometry, we take an overall observational
uncertainty of 10 per cent.

\subsubsection{Spectral fits}

\begin{figure}
\begin{center}
\includegraphics[width=0.62\textwidth,angle=90]{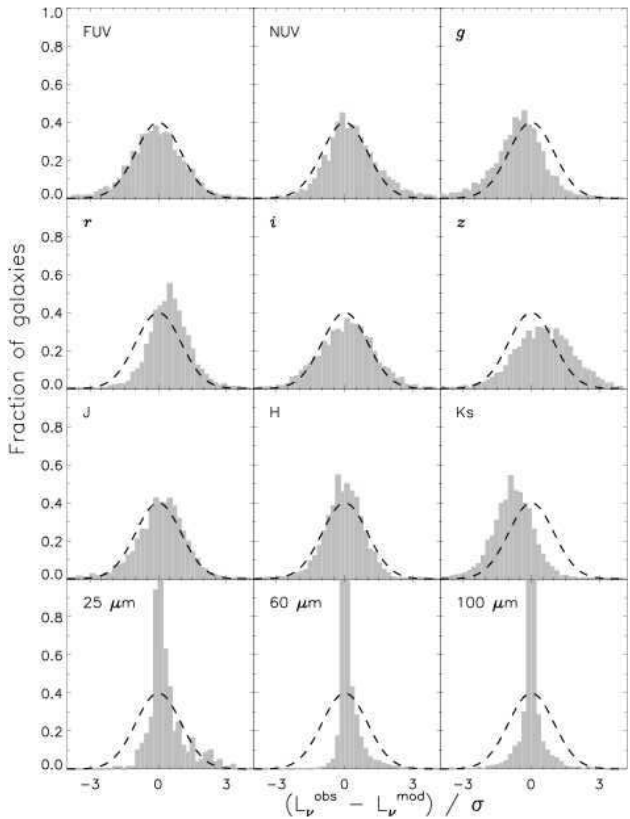}
\caption{Distribution of the difference between observed
luminosity $L_\nu^{\mathrm{obs}}$ and best-fit model luminosity
$L_\nu^{\mathrm{mod}}$, in units of the observational error
$\sigma$, for the galaxies in the matched \galex-SDSS-2MASS-\iras\
sample described in Section~\ref{dust:sample}. Each panel refers
to a different photometric band, as indicated. For reference,
the black dashed curve shows a Gaussian distribution with unit
standard deviation. The best-fit model for each galaxy was 
selected by fitting as many fluxes as available in the following
bands: {\it GALEX} ({\it FUV} and {\it NUV}), SDSS ($ugriz$),
2MASS ($JHK_s$), and \iras\ (12, 25, 60 and 100~\mic). 
}
\label{fig:dust3}
\end{center}
\end{figure}

We perform spectral fits by comparing the observed spectral
energy distribution of a galaxy to every model in the library
at the corresponding redshift. Specifically, for
each observed galaxy, we compute the $\chi^2$ goodness of fit
of each model. A model is characterised by a set of randomly 
drawn physical parameters. We build the likelihood distribution of
any given physical parameter for the observed galaxy by weighting
the value of that parameter in each model by the probability $\exp
(-\chi^2/2)$. We take our final estimate of the parameter to be
the median of the likelihood distribution, and the associated 
confidence interval to be the 16th--84th percentile range.

We use this approach to derive the likelihood distributions of
several physical parameters of the galaxies in our sample, based
on fits of the \galex\ {\it FUV} and {\it NUV}, SDSS $ugriz$,
2MASS $JHK_s$ and \iras\ 12-, 25-, 60- and 100-\mic\ fluxes. We
focus particularly on: the star formation rate averaged over the 
last $10^8$~yr, \sfr; the stellar mass, $M_\ast$; the specific 
star formation rate, $\ssfr=\sfr/M_\ast$; the dust mass, \mdust;
the total luminosity of the dust, \ldust, and the fraction of this
contributed by the diffuse ISM, \fmu.

We first check how well the model can reproduce the observed
spectral energy distributions of the galaxies in our sample. The
histograms in Fig.~\ref{fig:dust3} show, for each photometric band,
the distribution of the difference between the observed luminosity
$L_\nu^\mathrm{\,obs}$ and the best-fit model luminosity
$L_\nu^\mathrm{\,mod}$, in units of observational error $\sigma$.
Overall, the model provides remarkably consistent fits to the observed
ultraviolet, optical and infrared luminosities of the galaxies.
Fig.~\ref{fig:dust3} shows small systematic offsets in the $g$,
$r$ and $z$ bands, corresponding to an overestimate of the $g$-band
flux and underestimates of the $r$- and $z$-band fluxes of the order 
of $0.01$~mag. These offsets may originate from a deficiency in 
stellar population synthesis models, further worsened by the potential
contamination of the $g$ and $r$ bands by minor emission lines, which
we neglected in our corrections of Section~\ref{corrections}. We note
that the magnitude of these offsets is of the order of the uncertainties
in the SDSS AB calibration\footnote{see 
http://www.sdss.org/DR7/algorithms/fluxcal.html\#sdss2ab}.
In the near-infrared, there is a tendency
for the model to slightly overestimate the observed $K_s$-band luminosity
(by an amount corresponding to about $0.10$~mag). This offset is well
within the uncertainties of current stellar population synthesis models 
for this spectral region. We have checked that these offsets have
a negligible influence on our results.

\begin{figure}
\begin{center}
\includegraphics[width=0.57\textwidth,angle=90]{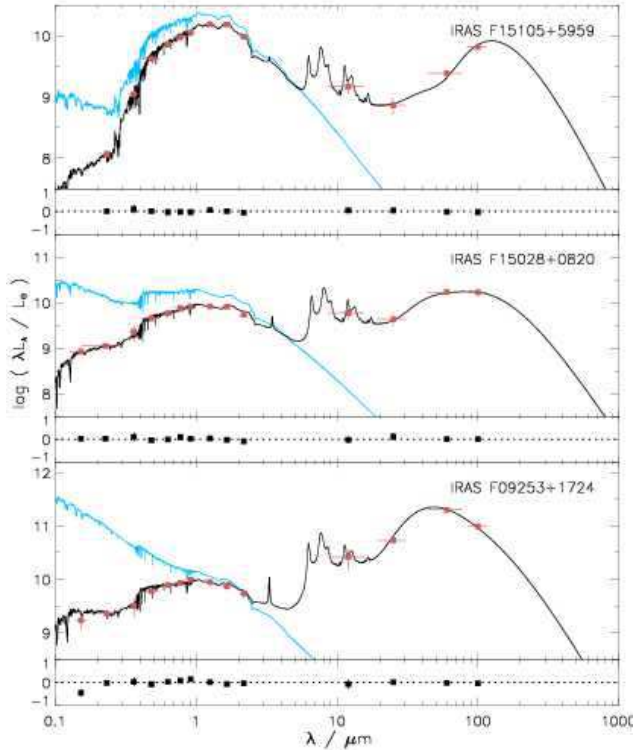}
\caption{Best-fit models (in black) to the observed spectral energy
distributions (in red) of 3 galaxies spanning wide ranges in
star formation and dust properties in the matched \galex-SDSS-2MASS-\iras\
sample described in Section~\ref{dust:sample}: from quiescent,
moderately dusty (top panel) to actively star-forming, highly dusty
(bottom panel). In each panel, the blue line shows the unattenuated
stellar spectrum. The red squares show the observed broadband 
luminosities (\galex\ {\it FUV}, {\it NUV}; SDSS $ugriz$; 2MASS $JHK_s$; 
\iras\ 12, 25, 60 and 100 \mic) with their errors as vertical
bars and the widths of the filters as horizontal bars. The fit
residuals $(L_\lambda^\mathrm{obs}-L_\lambda^\mathrm{mod})/
L_\lambda^\mathrm{obs}$ are shown at the bottom of each panel.}
\label{fig:dust4}
\end{center}
\end{figure}

To illustrate the quality of these fits, we show in Fig.~\ref{fig:dust4}
three examples of the best-fit spectral energy distributions of galaxies
spanning wide ranges in star formation and dust properties in our sample,
from quiescent, moderately dusty (IRAS F15105+5959, top panel) to actively
star-forming, highly dusty (IRAS F09253+1724, bottom panel). The middle
panel shows the spectral energy distribution of a galaxy corresponding 
roughly to the median star formation and dust properties of the sample
(IRAS F15028+0820).

%*****************************************************************************

\section[The relation between star formation and dust content]{The relation between star formation activity and dust content in galaxies}
\label{dust:rel_sfr_dust}

The method described in Section~\ref{dust:model} above allows
us to derive statistical constraints on the star formation rate 
averaged over the last $10^8$~yr, \sfr, the stellar mass, \mstar, 
the total dust luminosity, \ldust, the fraction of this contributed
by dust in the diffuse interstellar medium, \fmu, and the total
dust mass, \mdust, for the 3258 galaxies in the matched 
\galex-SDSS-2MASS-\iras\ sample described in Section~\ref{dust:sample}.

\begin{figure}
\begin{center}
\includegraphics[width=0.45\textwidth,angle=90]{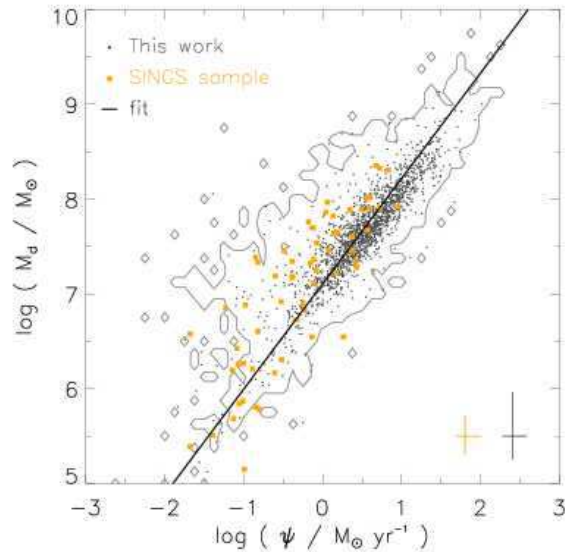}
\caption{Median-likelihood estimate of the dust mass, \mdust,
versus that of the star formation rate averaged over the last $10^8$~yr,
\sfr, for the matched \galex-SDSS-2MASS-\iras\ sample described
in Section~\ref{dust:sample} (in grey) and the SINGS galaxies (in
orange). The grey contour shows the distribution of the full 
matched sample of 3258 galaxies, while grey points show the 
distribution of the sub-sample of 1658 galaxies with highest-S/N
photometry in all bands (see text). The median error bars for
each sample are indicated in the lower-right corner. These correspond
to the median 16th--84th percentile range of the likelihood distributions
in \mdust\ and \sfr. The black line shows a linear fit to the grey 
points, computed as the bisector of the least-square regression lines
on each axis.} \label{fig:dust5}
\end{center}
\end{figure}

In Fig.~\ref{fig:dust5}, we show the resulting relation between dust
mass \mdust\ and star formation rate \sfr. The grey contour shows
the distribution for the full sample, while grey points show the 
distribution of a sub-sample of 1658 galaxies with highest-S/N
photometry at ultraviolet, optical and infrared wavelengths. This
high-S/N subsample includes only galaxies with relative photometric
errors less than 2$\sigma$ larger than the sample mean in each band. Also 
shown in the lower-right corner of Fig.~\ref{fig:dust5} are the median
errors in \sfr\ and \mdust, derived from the likelihood distributions
of these parameters for all the galaxies in the sample.

The correlation between dust mass and star formation rate is remarkably
tight in Fig.~\ref{fig:dust5}, spanning 4 orders of magnitude in both
\sfr\ and \mdust. We perform a linear fit to the grey points (i.e.,
the highest-S/N subsample) by computing the bisector of two least-square
regression lines \citep{Isobe1990}: one in \mdust\ as a function of \sfr,
and one in \sfr\ as a function of \mdust. This gives (black line in 
Fig.~\ref{fig:dust5})
\begin{equation}\label{equ1}
M_\mathrm{d} = (1.28\pm 0.02)\times10^7 \, 
(\sfr / \msun\ \mathrm{yr}^{-1})^{1.11 \pm 0.01}\,\msun\,.
\end{equation}
The uncertainties quoted in this expression are formal ones
derived by taking into account the confidence range in each
\sfr\ and \mdust\ measurement. These uncertainties are much smaller
than the intrinsic scatter of the relation (about 0.5~dex in
\mdust). Equation~(\ref{equ1}) can thus be used to estimate the
`typical' dust mass in a galaxy, based on the star formation rate.
To our knowledge it is the first time that such an expression is
calibrated for a large sample of galaxies.

We have investigated the extent to which the SDSS, \galex\ and \iras\ 
selections of our sample may introduce a bias in the relation between
star formation rate and dust mass derived from Fig.~\ref{fig:dust5}.
We used for this the library of stochastic models described in 
Section~\ref{library}. Since these models are normalised to total stellar
mass \citep{daCunha2008}, we assigned a random stellar mass (and scaled
dust mass and star formation rate) to each model in the library. We
drew the stellar masses uniformly in $\log(\mstar/\msun)$ between 8.5 
and 11.5, to be consistent with the distribution of galaxy stellar masses
in our sample. For each model in the library, we computed the expected
\galex, SDSS and \iras\ magnitudes in different redshift bins from $z=0$ 
to $z=0.20$. Then, we applied the same selection criteria as used for our
observed sample to the model library. These selection criteria introduce 
minimum detectable stellar mass, dust mass and star formation rate. We find
that, for example, at the low-redshift end of our sample ($z=0.0025$), the 
\iras\ flux limit tends to exclude galaxies with $\log(\mdust/\msun) < 4.5$. At
the typical redshift of our sample, $z=0.05$, the combination of
SDSS and \iras\ selections sets a minimum detectable dust mass of about
$10^7$~\msun\ and a minimum star formation rate of about $0.01$~\msun~yr$^{-1}$
(it also introduces serious incompleteness for galaxies with star formation 
rates less than 1~\msun~yr$^{-1}$). We have checked that these selection effects
do not affect significantly the empirical relation between the star formation 
rate and dust mass derived from Fig.~\ref{fig:dust5}, which is dominated by more
massive galaxies (including only galaxies with $\psi>1\,$\msun~yr$^{-1}$ and 
$\mdust>10^7\msun$ leaves the derived slope unchanged in equation~\ref{equ1}).

In Fig.~\ref{fig:dust5}, we also plot for comparison the dust masses
and star formation rates derived by \citet{daCunha2008} for the SINGS
galaxies (orange symbols).
These nearby galaxies follow the same 
relation as the galaxies in the matched \galex-SDSS-2MASS-\iras\ sample
studied here and extend to slightly lower dust masses and star formation
rates\footnote{\citet{daCunha2008} have checked that the dust masses
of the SINGS galaxies derived using the model described in 
Section~\ref{dust:model} above are typically within 50 per cent of those
estimated by \citet{Draine2007b} using a more sophisticated physical dust model.}.
We note that, at fixed star formation rate, the SINGS galaxies tend 
to have slightly larger dust masses than the galaxies in our sample. This
is likely to result from the different selection criteria of the SINGS sample.
The typical error bars in \mdust\ and \sfr\ are smaller for the 
SINGS galaxies than for the sample studied here, because a wider 
collection of observational constraints (especially in the mid-infrared)
were available to \citet{daCunha2008}.

In an attempt to understand which observations set the main constraints
on dust mass, we have investigated how the median-likelihood estimates 
of \mdust\ correlate with a wide range of galaxy colours. We find that \mdust\
correlates most strongly with the $F_\nu^{\,100}/F_\nu^{\,g}$ colour, where 
$F_\nu^{\,100}$ is the flux density in the \iras\ 100-\mic\ band and $F_\nu^{\,g}$
that in the SDSS $g$ band. The Spearman rank coefficient for this correlation
is $r_S=0.56$, indicating a positive correlation at the $23\sigma$
level for this sample size (we note that the correlation is not improved by the
inclusion of an ultraviolet band such as \galex\ FUV). The reason for this 
correlation is that $F_\nu^{\,100}/F_\nu^{\,g}$ traces the primary contributor
to \mdust: cold dust in the diffuse ISM ($M_\mathrm{C}^{\,\mathrm{ISM}}$ in 
eq.~\ref{eq:mdust}). This component is also mainly responsible for the attenuation
of the emission from stars older than $10^7$~yr, which dominate the SDSS $g$-band light,
and it is the main contributor to the emission at 100~\mic\ (a more detailed discussion
of the set of observables necessary to constrain \mdust\ can be found in Section
3.2.2 of \citealt{daCunha2008}).

\begin{figure*}
\begin{center}
\includegraphics[width=1.\textwidth]{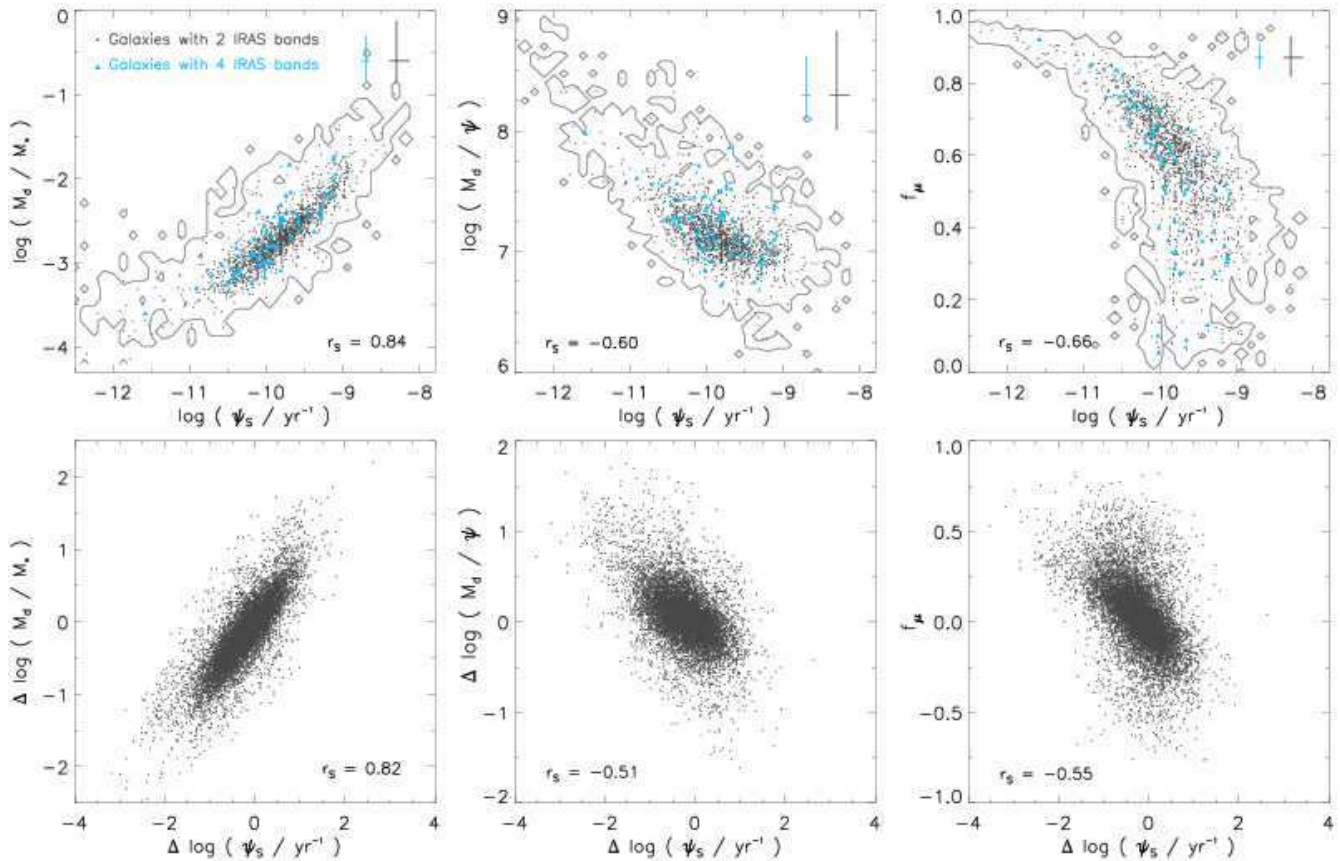}
\caption{{\it Top:} Median-likelihood estimates of 3 galaxy
properties against specific star formation rate, \ssfr. {\it Left panel}:
ratio of dust mass to stellar mass, $\mdust/\mstar$. {\it Middle
panel}: ratio of dust mass to star formation rate, $\mdust/\sfr$
(which may be used as a proxy for the dust-to-gas ratio; see 
Section~\ref{dust:rel_sfr_dust}). {\it Right panel}: fraction of total infrared
luminosity contributed by dust in the ambient ISM, $f_\mu$. In each
panel, the grey contour shows the distribution of the full matched 
\galex-SDSS-2MASS-\iras\ sample described in Section~\ref{dust:sample},
while the points show the distribution of the sub-sample of 1658 
galaxies with highest-S/N photometry in all bands. The blue triangles
show galaxies with complete \iras\ information in the 12-, 25-, 60- 
and 100-\mic\ bands. The error bars represent the median confidence
ranges in each parameter. {\it Bottom:} differences in the same
quantities as in the top panels for pairs of galaxies closely matched in
stellar mass (see text for detail). The Spearman rank coefficient $r_{S}$
is indicated in each panel.}\label{fig:dust6}
\end{center}
\end{figure*}

In Fig.~\ref{fig:dust6}, we explore the relations between specific 
star formation rate \ssfr\ and three particularly interesting physical 
properties of the galaxies in our sample: the dust-to-stellar mass ratio
$\mdust/\mstar$, the ratio of dust mass to star formation
rate $\mdust/\sfr$, and the fraction $f_\mu$ of the total infrared
luminosity \ldust\ contributed by dust in the diffuse ISM. As in 
Fig.~\ref{fig:dust5}, grey contours show the relations for the full
sample of 3258 galaxies, while individual points show the relations for
the 1658 galaxies with highest-S/N photometry. The top panels of 
Fig.~\ref{fig:dust6} indicate that $\mdust/\mstar$, 
$\mdust/\sfr$ and $f_\mu$ are all strongly correlated with \ssfr:
the Spearman rank coefficients for the full sample are $r_{S} = 0.84$, 
$-0.60$ and $-0.66$, respectively, indicating that the correlations are
significant at more than $20\sigma$ level for this sample size 
(similar results are obtained when using only the high-S/N subsample).
It is important to check that stellar mass is not the main driver for
these strong correlations. To verify this, in the bottom panels of 
Fig.~\ref{fig:dust6}, we show differences in the same
quantities as in the top panels for pairs of galaxies closely matched
in stellar mass (we have included all possible galaxy pairs for any 
stellar mass \mstar\ in the full sample). Specifically, for each galaxy
pair, we plot the difference in specific star formation rate between
the two galaxies, [$\Delta \log (\ssfr/ \mathrm{yr}^{-1})$], 
against the difference in dust-mass to stellar-mass ratio, [$\Delta 
\log(\mdust/\mstar)$], the difference in ratio of dust mass to 
star formation rate, [$\Delta \log (\mdust/ \sfr )$], and the 
difference in fraction of total infrared luminosity contributed by 
the diffuse ISM [$\Delta f_\mu$]. The fact that the strong correlations 
subsist from the top to the bottom panels of Fig.~\ref{fig:dust6} 
demonstrates that stellar mass is not the main driver of these correlations.

\begin{figure*}
\begin{center}
\includegraphics[width=0.95\textwidth]{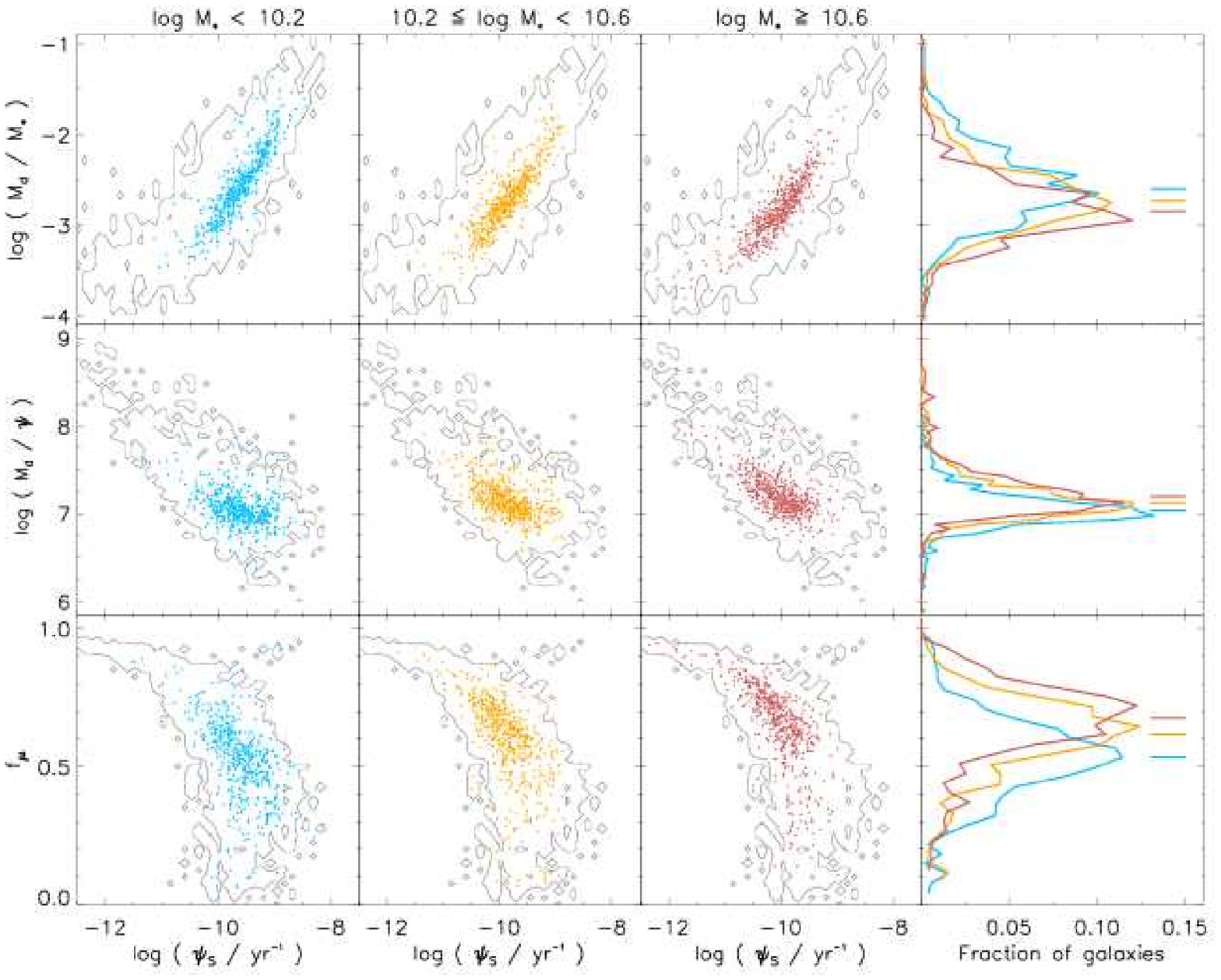}
\caption{Same relations as in the top panels of Fig.~\ref{fig:dust6}
plotted in 3 stellar-mass ranges: $\log(\mstar/\msun) <10.2$ (blue); 
$10.2 \leq \log(\mstar/\msun)< 10.6$ (orange); $\log(\mstar/\msun) \geq
10.6$ (red). The right-hand panels show the corresponding distributions
of the dust-to-stellar mass ratio $\mdust/\mstar$, the ratio of dust
mass to star formation rate $\mdust/\sfr$ and the fraction 
$f_\mu$ of total infrared luminosity contributed by dust in the ambient
ISM, along with their median values. The contours are the same as
in the top panels of Fig.~\ref{fig:dust6}.}
\label{fig:dust7}
\end{center}
\end{figure*}

It is of interest to check the extent to which the properties of galaxies
in Fig.~\ref{fig:dust6} actually depend on stellar mass. Several
studies have shown that the star formation activity of a galaxy tends to
decrease with increasing stellar mass (e.g., \citealt{Brinchmann2004}). 
In Fig.~\ref{fig:dust7}, we plot $\mdust/\mstar$, $\mdust/\sfr$
and $f_\mu$ as a function of \ssfr\ for 3 different 
stellar-mass ranges chosen to contain roughly similar numbers of
galaxies. The distributions in the various quantities on
the $y$-axis and their median values are displayed in the right-hand
panels. Fig.~\ref{fig:dust7} shows that the relation between $\mdust/
\mstar$ and \ssfr\ depends little on stellar mass, with only 
a slight tendency for the less massive galaxies to have somewhat 
higher $\mdust/\mstar$. Also, the quantity $\mdust/\sfr$, which 
traces the dust-to-gas ratio, tends to 
increase slightly with stellar mass. The quantity displaying the 
strongest dependence on stellar mass in Fig.~\ref{fig:dust7} is the
fraction $f_\mu$ of the total infrared luminosity contributed by dust
in the diffuse ISM: the median value of this parameter increases from
0.5 for low-mass galaxies to 0.7 for high-mass ones. We note that
the 3 stellar-mass bins considered here span only a small 
dynamic range in \mstar, which may not be enough to reveal the full
dependence of the considered properties on stellar mass. It is tempting
to interpret the relations of Figs.~\ref{fig:dust6} and \ref{fig:dust7}
as evolutionary sequences, where the build-up of stellar mass would be
accompanied by gas consumption and dust enrichment. We return to this
point Section~\ref{dust:che_models} below.

We have also checked that the non-detection in the 12- and 25-\mic\ 
\iras\ bands of most galaxies in the sample (Section~\ref{fir_phot})
has a negligible influence on the results of Fig.~\ref{fig:dust6}. This
is illustrated by the blue triangles in the top panels of the figure,
which mark galaxies with complete \iras\ information in the 12-, 25-,
60- and 100-\mic\ bands. As expected, the median errors in the 
derived parameters of these galaxies are typically smaller than 
those for the full sample (as indicated in the upper-right corner of 
each diagram). Fig.~\ref{fig:dust6} shows that galaxies with complete
\iras\ information follow the same trends as the rest of the sample.
Thus, the availability of only the 60- and 100-\mic\ \iras\ flux 
densities for most galaxies in the sample does not appear to 
systematically bias the results.

The strength of the correlations between \ssfr\ and $\mdust/\mstar$,
$\mdust/\sfr$ and $f_\mu$ in Fig.~\ref{fig:dust6} suggests that
the specific star formation rate is a fundamental diagnostic of 
the ISM properties in galaxies. The left-hand panels of 
Fig.~\ref{fig:dust6} show that, for example, the galaxies with 
highest specific star formation rate are also the most dust-rich. The
reason for this can be best understood with the help of the middle panels.
In these diagrams, $\mdust/\sfr$ may be regarded as a proxy for
the dust-to-gas ratio, since \sfr\ is tightly related to the gas mass
(\hi\ + H$_2$) by virtue of the Schmidt-Kennicutt law \citep{Schmidt1959,
Kennicutt1998b}.  
The inverse correlation between $\mdust/\sfr$ and \ssfr\ suggests that
the dust-to-gas ratio is larger in galaxies with low specific star 
formation rate, which have presumably exhausted their gas reservoir. 
In contrast, galaxies with high specific star formation rate and hence 
large gas reservoirs tend to have lower dust-to-gas ratio [we note that
supernovae and young asymptotic-giant-branch (AGB) stars will produce 
large amounts of new dust in these galaxies; see e.g. \citealt{Dwek1998}].
The left-hand panels of Fig.~\ref{fig:dust6} show that, as a result, a
young, actively star-forming galaxy with low dust-to-gas ratio may still 
be highly dusty (in the sense of a high $\mdust/\mstar$) because it
contains large amounts of interstellar gas (see also 
Section~\ref{dust:che_models} below).
Finally, the right-hand panels of Fig.~\ref{fig:dust6} show that stellar
birth clouds are the main contributors to dust heating in actively 
star-forming galaxies, while in more quiescent galaxies, the bulk of dust
emission arises from the heating of dust by older stars in the diffuse 
interstellar medium. This trend was previously noted for SINGS galaxies by
\citet{daCunha2008}. 

\begin{figure}
\begin{center}
\includegraphics[width=0.47\textwidth]{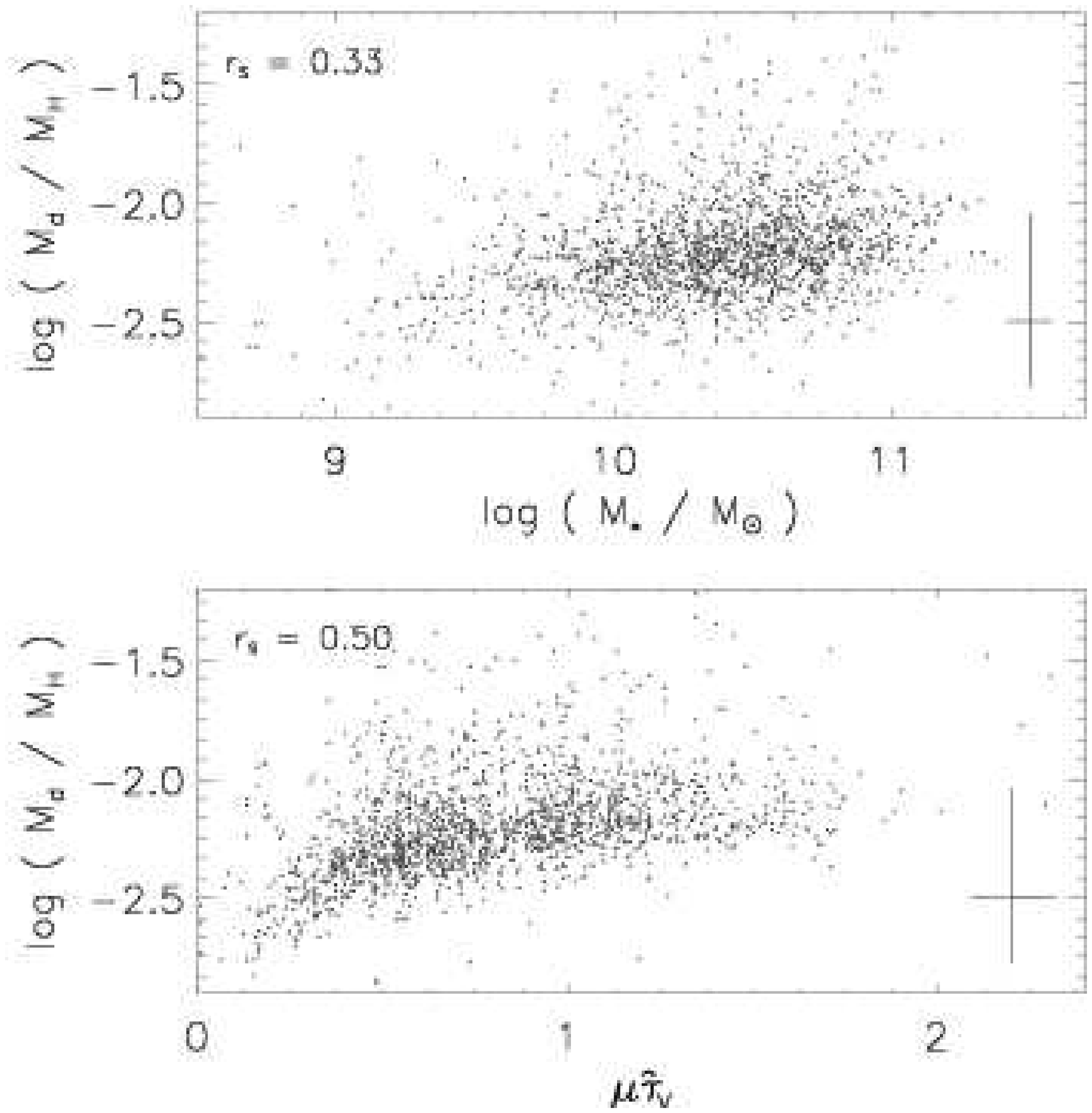}
\caption{{\it Top}: ratio of dust mass \mdust\ to gas mass,
$\mgas= M_\mathrm{H\,\scriptsize\textsc{i}}+M_{\mathrm{H}_2}$, 
plotted against the stellar mass \mstar\ for the 1658 galaxies with
highest-S/N photometry in the matched \galex-SDSS-2MASS-\iras\ sample
described in Section~\ref{dust:sample}. {\it Bottom}: $\mdust/\mgas$
ratio plotted against the dust attenuation optical depth in the
diffuse ISM, $\mu\tauv$, for the same sample. The median error bars
are indicated in the lower-right corner of each panel.}
\label{fig:dust8}
\end{center}
\end{figure}

We may attempt to combine the above estimates of the dust mass \mdust\ 
for the galaxies in our sample with gas-mass estimates derived
from the star formation rate \sfr\ via the Schmidt-Kennicutt law,
to constrain the dust-to-gas ratio. The Schmidt-Kennicutt takes the
form \citep{Kennicutt1998}
\begin{equation}
\mathrm{\Sigma}_\mathrm{SFR}= (2.5\pm0.7) \times 10^{-4} \, 
\mathrm{\Sigma}^{1.4\pm0.15}_\mathrm{H}\,,
\label{eq:kennicutt}
\end{equation}
where $\mathrm{\Sigma}_\mathrm{SFR}$ is the surface density of 
star formation, expressed in $\msun\,\mathrm{yr}^{-1}\,
\mathrm{kpc}^{-2}$, and $\mathrm{\Sigma}_\mathrm{H}$ is the 
surface mass density of \hi+H$_2$ gas, expressed in $\msun
\,\mathrm{pc}^{-2}$. This formula was derived using a 
\citet{Salpeter1955} IMF with lower and upper cut-offs $0.1$ and
$100\, \msun$. We adjust the scaling coefficient by a factor of
1.6 down to account for the fact that the \citet{daCunha2008} 
model used here relies on a \cite{Chabrier2003} IMF with same
cut-offs. We compute $\mathrm{\Sigma}_\mathrm{SFR}$ for the 
galaxies in the high-S/N subsample studied above by dividing 
our estimates of \sfr\ by the area defined by the minor and major 
axes of the $r$-band isophote at 25\,mag\,arcsec$^{-2}$ (see 
Section~\ref{dust:inclination}). We then compute 
$\mathrm{\Sigma}_\mathrm{H}$ and hence \mgas\ using 
equation~(\ref{eq:kennicutt}).

In Fig.~\ref{fig:dust8}, we plot the dust-to-gas ratio $\mdust/\mgas$
derived in this way against the stellar mass \mstar\ and the dust 
attenuation optical depth in the diffuse ISM $\mu\tauv$ for the 
galaxies in the high-S/N subsample. The median errors
in $\mdust/\mgas$ are quite large. Yet, this ratio appears to
correlate with both \mstar\  ($r_S=0.33$; $13\sigma$ significance
level) and $\mu\tauv$ ($r_S=0.50$; $20\sigma$ significance level).
The correlation with \mstar\ is consistent with the tight correlation
between stellar mass and gas-phase metallicity for SDSS star-forming
galaxies \citep[e.g.][]{Tremonti2004}. In fact, the strong correlation 
between dust-to-gas ratio and gas-phase metallicity has been noted
in several previous studies (e.g. \citealt{Issa1990,Schmidt1993,
Lisenfeld1998}). We note that the relation between $\mdust/\mgas$
and $\mu\tauv$ in Fig.~\ref{fig:dust8} has too much scatter for 
the dust optical depth in the diffuse ISM to be used as a reliable 
proxy for the dust-to-gas ratio in galaxies.

%****************************************************************

\section{Discussion}\label{dust:discussion}

In this section, we briefly mention model uncertainties and discuss
potential biases introduced by inclination
effects and the presence of obscured AGNs in the results derived in the
previous sections. We also compare these results with the predictions of
models of chemical and dust evolution to illustrate the implications
of our study for the evolution of star-forming galaxies.

\subsection{Model uncertainties} \label{dust:uncertainties}

The \citet{daCunha2008} model used in Section~\ref{dust:model} 
to derive the physical properties of galaxies in the matched 
\galex-SDSS-2MASS-\iras\ sample relies on a combination of the
latest version of the \citet{Bruzual2003} population synthesis
code and the simple two-component dust model of \citet{Charlot2000}.
Recently, \citet{Conroy2008} have investigated the uncertainties
inherent to such models, in particular, those arising from poorly
understood phases of stellar evolution (blue horizontal branch stars,
blue straggler stars, thermally pulsing AGB stars), the IMF and the
properties of dust in the interstellar medium.\footnote{We note that
\citet{Conroy2008} inappropriately refer to the \citet{Charlot2000} model
as a `uniform screen of dust'. In reality, by design, the effective 
attenuation curve in the \citet{Charlot2000} model reflects the probability
density for the absorption of photons emitted in all directions by stars
in all locations within a galaxy (see their eq.~[18]). The shape of this
curve -- which is the quantity primarily constrained by observations -- is controlled by 
the combination of the optical properties and the spatial distribution of
the dust. Allowing for variations in this shape therefore allows one
to explore variations in the spatial distribution of the dust in a galaxy
(as exemplified in section 4 of \citealt{Charlot2000}).} Uncertainties
linked to the IMF will affect our analysis in the same way as pointed out
by these authors. In our case, the uncertainties linked to stellar
evolution are somewhat minimized by the fact that blue-straggler and
horizontal-branch stars have a negligible contribution to the emission
from star-forming galaxies. Also the version of the \citet{Bruzual2003} 
models used here rely on the improved, empirically calibrated models of 
\citet{Marigo2007} to better account for the contribution by thermally 
pulsing AGB stars to the near-infrared emission from galaxies
\citep[see][]{Bruzual2007}. We refer the reader to the original studies of
\citet{Charlot2000} and \citet{daCunha2008} for discussions of the
uncertainties in the dust model (e.g., spatial distribution and optical
properties of dust grains; typical lifetime of giant molecular clouds)
and their influence on the ultraviolet and infrared emission from galaxies.

\subsection{The effect of inclination} \label{dust:inclination}

\begin{figure}
\begin{center}
\includegraphics[width=0.69\textwidth,angle=90]{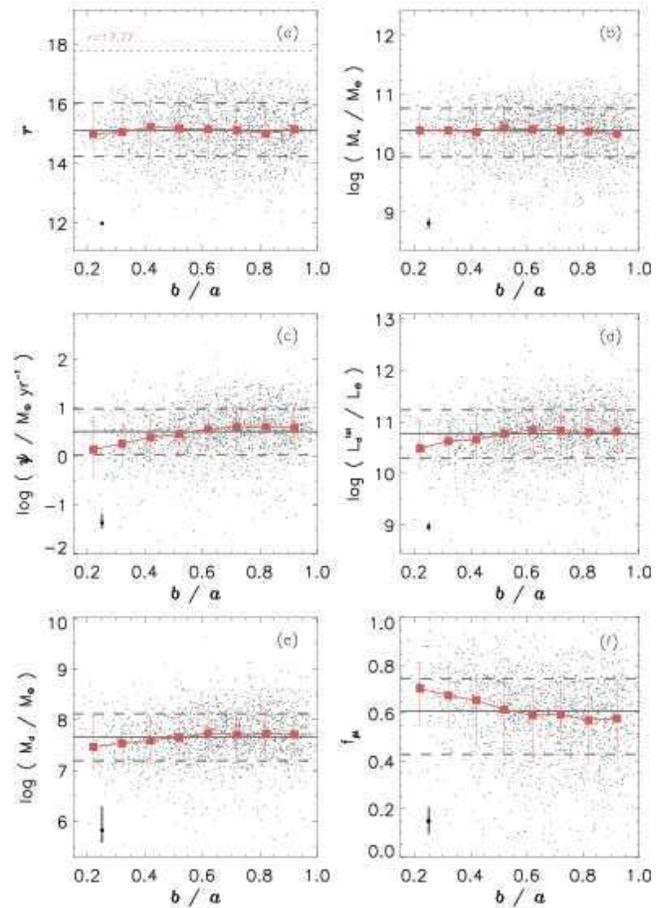}
\caption{Galaxy properties plotted against ratio of minor
to major axes $b/a$ (used as a proxy for inclination) for the 
1658 galaxies with highest-S/N photometry in the matched
\galex-SDSS-2MASS-\iras\ sample
described in Section~\ref{dust:sample}: (a) apparent $r$-band
magnitude, $r$; (b) stellar mass, \mstar; (c) star formation rate
averaged over the last $10^8$~yr, \sfr; (d) total infrared luminosity,
\ldust; (e) dust mass, \mdust; (f) fraction of \ldust\ contributed by
dust in the diffuse ISM, \fmu. Red squares and error bars indicate 
the median values and 16th--84th percentile ranges in bins of $b/a$.
For reference, the horizontal solid and dashed grey lines indicate 
the median value and 16th--84th percentile range for the whole sample
(i.e. including all inclinations). The median observational error
in $r$ is indicated in the lower-left corner of panel (a). In each
panel from (b) to (f), the median confidence interval in the 
likelihood estimates of the considered parameter is indicated in 
the lower-left corner.} \label{fig:dust9} 
\end{center}
\end{figure}

The model used in Section~\ref{dust:model} to derive the physical 
properties of galaxies in the matched \galex-SDSS-2MASS-\iras\ sample 
is limited to angle-averaged spectral properties. However, the observed
fluxes and colours of galaxies are observed to depend on inclination,
especially for spiral galaxies (e.g., \citealt{Maller2009}). SDSS 
galaxies have been morphologically classified using the concentration
index $C$, defined as the ratio of the radii enclosing 90 per cent and
50 per cent of the $r$-band luminosity (e.g.~\citealt{Shimasaku2001,Strateva2001}).
We find that 64 per cent of 
the galaxies in our sample have concentration indices typical of 
late-type galaxies (i.e., $C<2.6$). Thus, we must check that inclination
effects do not introduce any strong bias in the results of 
Section~\ref{dust:rel_sfr_dust}. 

We use the ratio $b/a$ of minor to major axes of the SDSS $r$-band 
isophote at 25\,mag\,arcsec$^{-2}$ as a proxy for disc inclination. This
is justified by the way in which galaxy ultraviolet and optical colours 
correlate with $b/a$. For example, we find that colours such as $g-r$ and 
$NUV-r$ correlate significantly with $b/a$, with the smaller $b/a$ 
(presumably tracing higher disc inclinations) corresponding
to the redder ultraviolet/optical colours.
 In Fig.~\ref{fig:dust9}, we plot several properties of the galaxies 
in the high-S/N subsample (Section~\ref{dust:rel_sfr_dust}) against the 
axis ratio $b/a$: apparent $r$-band magnitude, $r$; stellar mass,
\mstar; star formation rate averaged over the last $10^8$~yr, \sfr;
total infrared luminosity\footnote{We note that the values of \ldust 
derived from our multi-wavelength analysis are typically 0.2~dex larger 
than those computed using the simple empirical approximation of
eqs.~(\ref{Helou1988_1})--(\ref{Helou1988_2}).}, \ldust;  dust mass,
\mdust; and fraction of \ldust\ contributed by dust in the diffuse ISM,
\fmu.

We first highlight a selection effect: at low $b/a$ (i.e. high
inclination), the sample includes almost no galaxy fainter than
$r \sim 16$\,mag (Fig.~\ref{fig:dust9}a). This is because of the
combined requirements of infrared detection by \iras\ and 
ultraviolet detection by {\it GALEX}: edge-on galaxies in our 
sample must be optically bright enough that the limited relative 
dust content required for some ultraviolet photons to escape in 
the plane of the disc be also sufficient to warrant detection in 
the infrared. It is important to note that our restriction to the
high-S/N subsample in Fig.~\ref{fig:dust9}a explains the lack of 
galaxies at the faintest magnitudes near the $r=17.77$ selection
cut-off even at low inclinations.

Fig.~\ref{fig:dust9}b shows that the median-likelihood estimates
of stellar mass do not vary significantly with axis ratio, implying
that inclination has a negligible influence on \mstar\ constraints. 
This is not surprising, since constraints on \mstar\ are dominated by the
near-infrared emission, which is little sensitive to dust attenuation.
Estimates of the star formation rate exhibit a stronger dependence 
on inclination (Fig.~\ref{fig:dust9}c): the typical \sfr\ of face-on
galaxies ($b/a \approx 1$) is a factor of about 3 larger than that
of edge-on galaxies. This difference is significant when compared to
the median confidence interval in \sfr\ estimates (indicated on the 
figure). It is consistent with the selection bias identified in
Fig.~\ref{fig:dust9}a against very dusty edge-on galaxies. This 
effect is also responsible for the lower dust masses and dust 
luminosities derived for edge-on compared to face-one galaxies in 
Figs.~\ref{fig:dust9}d and \ref{fig:dust9}e. Finally, 
Fig.~\ref{fig:dust9}f shows the contribution \fmu\ to the total
infrared luminosity by dust in the diffuse ISM is systematically
higher in edge-on galaxies than in face-on ones. This is likely
to result from the lower typical specific star formation rate 
of edge-on galaxies (Figs.~\ref{fig:dust9}b and \ref{fig:dust9}c;
see the correlation between \fmu\ and \ssfr\ in the right-most panels
of Fig.~\ref{fig:dust6}).

We emphasise that the effects identified above are weak compared to
the typical dispersion in the estimated parameters. Thus, inclination
effects can account only for a minor fraction of the intrinsic scatter
in the relations studies in Section~\ref{dust:rel_sfr_dust}.

\subsection{Contamination by AGN hosts} \label{dust:AGN}

In Section~\ref{dust:model}, we have interpreted the spectral energy 
distributions of galaxies in the matched \galex-SDSS-2MASS-\iras\ sample
using the model of \cite{daCunha2008}. An important limitation of this 
model is that it does not include the potential contribution by an AGN 
to the infrared emission of a galaxy (e.g.~\citealt{DeGrijp1985}). For
this reason, in Section~\ref{dust:sample}, we have excluded from our 
sample all potential AGN hosts on the basis of their optical-line emission.
However, optically-thick AGNs would not be eliminated using this method.
This question is a fair concern, as the fraction of AGN hosts is known
to increase with the total infrared luminosity \citep{Veilleux1995,
Cao2006}. In addition, evidence for dust-enshrouded AGN activity has 
been found in some ULIRGs and LIRGs of the \iras\ catalogue 
(\citealt{Dudley1997,Imanishi2000}). In an infrared-selected sample of
high-redshift ($z\sim0.8$) galaxies with infrared luminosities similar to 
those of the galaxies in our sample, \citet{Elbaz2002} find $12\pm5$~per
cent (5/41) of AGN hosts based on available observations with {\it Chandra 
X-ray Observatory} \citep{Brandt2001,Hornschemeier2001}. It is conceivable,
therefore, that a similar fraction of galaxies in our sample host an 
obscured AGN.

Hot dust in the torus surrounding an obscured AGN is expected to radiate 
mostly in the mid-infrared. This should have only a minor influence on our
analysis, since most galaxies in our sample are not detected in the \iras\
12- and 25-\mic\ bands (Fig.~\ref{fig:dust6}). However, some studies 
suggest that AGNs may be also contribute significantly to the far-infrared
emission from galaxies. For example, in the analysis of a matched sample of
SDSS-DR2 and \iras\ galaxies, \citet{Pasquali2005} find that known AGN 
hosts exhibit a significant excess of 60- and 100-\mic\ emission,
amounting to a typical excess infrared luminosity of 0.18~dex. Yet, it
remains unclear whether this excess emission arises from (cold) dust heated
by the AGN or by a population of young stars not detected in the optical
SDSS spectra. In another recent study, \cite{Salim2009} analyse the 
24~\mic\ emission of a sample of optically-selected galaxies observed with
the {\it Spitzer Space Telescope} \citep{Werner2004} out to redshift
$z=1.4$. They conclude that AGNs identified in the X-rays and the optical
do not contribute significantly to the 24\,\mic\ emission in their sample.
Moreover, optically selected AGNs do not present any evidence for an 
excess infrared luminosity, while obscured AGNs detected in the X-rays can
contribute only up to 50~per cent of the total infrared luminosity.

We have checked that the possible excess infrared luminosity arising from 
an obscured AGN in these previous studies is within the typical confidence
range of \ldust\ estimates for the galaxies in our sample. We conclude 
that the potential contamination of the 60- and 100-\mic\ \iras\ flux
densities by obscured AGNs would not alter significantly our conclusions.

\subsection{Comparison with models of chemical and dust evolution} \label{dust:che_models}

\begin{figure*}
\begin{center}
\includegraphics[width=0.8\textwidth]{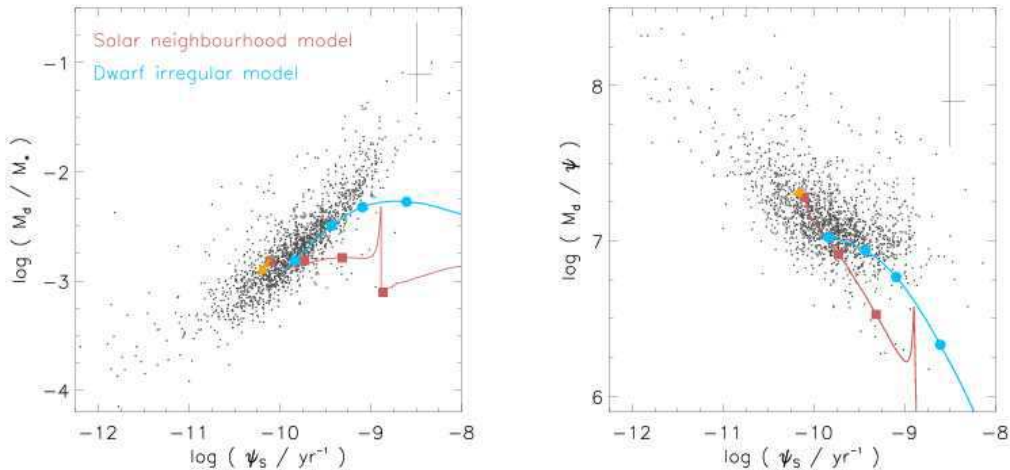}
\caption{Comparison of the physical properties derived in 
Section~\ref{dust:rel_sfr_dust} for the high-S/N galaxies in the matched 
\galex-SDSS-2MASS-\iras\ sample described in Section~\ref{dust:sample}
(grey points) with the theoretical predictions of the 
\protect\citet{Calura2008} models of chemical and dust evolution. 
{\it Left}: dust-to-stellar mass ratio $\mdust/\mstar$ as a 
function of specific star formation rate \ssfr. {\it Right}: ratio
of dust mass to star formation rate $\mdust/\sfr$ as a function
of \ssfr. The evolution of the solar-neighbourhood model of 
\protect\cite{Calura2008} is shown in red, and that of the 
dwarf-irregular galaxy model in blue. Symbols indicate the locations
of these models at ages 1, 3, 6 and 12\,Gyr. The orange arrow shows
the effect of adding a bulge component to the 12-Gyr old 
solar-neighbourhood model to mimic the Milky-Way properties (see text
for detail).}  
\label{fig:dust10}
\end{center}
\end{figure*}

In this section, we compare the physical properties derived in 
Section~\ref{dust:rel_sfr_dust} for the galaxies in the matched
\galex-SDSS-2MASS-\iras\ sample with the predictions of chemical
evolution models including the formation and destruction of dust
grains in the ISM. We focus on the recent models of the chemical
and dust evolution of galaxies in a wide range of star formation
histories by \cite{Calura2008}. These models, which are based on
the formalism developed by \cite{Dwek1998}, include the production
of dust in the cold envelopes of low- and intermediate-mass stars
during the AGB phase and in the expanding ejecta of type-II and
type-Ia supernovae. Dust grains are destroyed mainly by shock 
waves caused by supernovae explosions. The \cite{Calura2008} 
models allow one to compute the evolution of the stellar and 
dust masses of galaxies as a function of the star formation history.

We are particularly interested here in the \cite{Calura2008} models
for the solar neighbourhood and for a dwarf-irregular galaxy with 
continuous star formation. The solar-neighbourhood model aims at 
reproducing the properties of the Galactic-disc component in a 
2\,kpc-wide ring located at 8\,kpc from the Galactic centre. In 
this model, which assumes gas infall but no outflow, the Galactic
disc is formed in two main episodes. The exact expression of the 
star formation history, taken from \citet{Chiappini1997}, is 
essentially a function of the gas surface mass density with a 
critical threshold for star formation. In the model of dwarf-irregular
galaxy, the star formation rate is expressed as a Schmidt-Kennicutt
law and increases continuously with time. This model assumes gas infall
and galactic winds. We refer the reader to the original paper of
\citet{Calura2008} for more detail on these models.

In Fig.~\ref{fig:dust10}, we compare the median-likelihood
estimates of $\mdust/\mstar$ and $\mdust/\sfr$ as a
function of \ssfr\ for the galaxies in the high-S/N subsample
of Section~\ref{dust:rel_sfr_dust} with the predictions of the 
solar-neighbourhood model (red filled squares) and the 
dwarf-irregular galaxy model (blue filled circles) of \cite{Calura2008},
at ages between 1 and 12~Gyr. At late ages, both models coincide
remarkably well with the range of derived physical parameters of
the \galex-SDSS-2MASS-\iras\ galaxies. The solar-neighbourhood 
model of \cite{Calura2008} does not include any bulge component.
To locate the Milky Way in Fig.~\ref{fig:dust10}, we add to this
model a bulge stellar mass of $1/3$ that of the disc. We assume
that this bulge component has negligible contributions to \mdust\
and \ssfr\ (consistent with the chemical evolution model of a
12-Gyr old spheroidal galaxy by \citealt{Pipino2005}). The resulting
Milky-Way model has a total stellar mass $\mstar= 4
\times10^{10}$~\msun, a star formation rate $\sfr = 
2.3$~\msun\ yr$^{-1}$ and a dust mass $\mdust = 4.5\times10^{7}$~\msun\
(head of the orange arrow in Fig.~\ref{fig:dust10}). These parameters
are typical of moderately star-forming and moderately dusty galaxies
in our sample.

A valuable feature of the models shown in Fig.~\ref{fig:dust10} is
that they provide a framework to interpret the time evolution of the
ISM properties of galaxies. Both the solar-neighbourhood and the
dwarf-irregular galaxy models predict that, overall, $\mdust/\mstar$
should drop and $\mdust/\sfr$ should rise as the specific
star formation rate \ssfr\ declines. These trends, and the good general 
agreement of the models with the derived physical properties of the 
galaxies in our sample in Fig.~\ref{fig:dust10}, suggest that the 
relations identified between $\mdust/\mstar$, $\mdust/\sfr$,
$f_\mu$ and \ssfr\ in Section~\ref{dust:rel_sfr_dust} (Fig.~\ref{fig:dust6})
may be the result, at least in part, of an evolutionary sequence. 
In detail, the interpretation of this sequence depends on several
key assumptions of the chemical evolution models, such as the 
star formation history, the influence of gas infall and outflow,
and the dust production and destruction rates. In the 
\cite{Calura2008} models of Fig.~\ref{fig:dust10} (which both
include gas infall, the dwarf irregular model also including
gas outflow), a young galaxy has an ISM characterized
by large amounts of dust-poor gas and forms stars at a
very high rate. As the galaxy continues to consume gas into
star formation, the stellar mass rises and 
the star formation rate declines, causing \ssfr\ to drop.
The dust mass increases steeply during the first billion years
of evolution, when the star formation rate is highest and large
amounts of dust are produced in supernova ejecta and in the 
envelopes of AGB stars. As star formation drops, the dust mass
\mdust\ also drops, because the production of new dust grains
cannot balance any more their destruction in the ISM. This
general picture, which is consistent with our results, is also 
in broad agreement with the predictions of simple closed-box 
chemical evolution models (e.g., 
\citealt{Dwek2000,Edmunds2001}).

%******************************************************************************
\section{Summary and conclusion}\label{dust:conclusion}

In this paper, we have assembled a sample of 3258 local galaxies
with photometric observations from at ultraviolet, optical and 
infrared wavelengths. These galaxies were primarily selected from
the SDSS DR6 spectroscopic sample cross-correlated with \iras\ 
all-sky catalogues (which are flux-limited at 60\,\mic). The 
optical and infrared observations were supplemented with matched
ultraviolet and near-infrared data from \galex\ and 2MASS, 
respectively. The size of this sample represents a significant 
improvement over previous multi-wavelength studies of galaxies. 
We have used the model of \citet{daCunha2008} to interpret the observed 
spectral energy distributions of the galaxies in this sample in terms
of statistical constraints on the star formation rate, stellar mass,
dust attenuation, dust luminosity, fraction of dust luminosity 
contributed by the diffuse ISM, and dust mass.

We have focused on a subsample of 1658 galaxies with highest-S/N
photometry to investigate several significant correlations between
various derived physical properties of galaxies. In particular, we
find that the star formation rate averaged over the last $10^8$~yr,
\sfr, correlates remarkably well with galaxy dust mass \mdust\ over
4 orders of magnitude in both quantities. The simple empirical
recipe $\mdust = (1.28\pm 0.02)\times10^7 \, (\sfr / \msun\
\mathrm{yr}^{-1})^{1.11 \pm 0.01}\,\msun$ may be used to roughly 
estimate the total dust mass of a galaxy as a function of the star
formation rate. We also find that the dust-to-stellar mass ratio
$\mdust/\mstar$, the ratio of dust mass to star formation rate 
$\mdust/ \sfr$ and the fraction \fmu\ of dust luminosity \ldust\ 
contributed by the diffuse ISM correlate strongly with the specific
star formation rate \ssfr\ for the galaxies in the high-S/N subsample.
Some of these trends had already been anticipated in the analysis
of a much smaller sample of 66 SINGS galaxies by \citet{daCunha2008}.

To investigate the origin of these correlations between various
physical properties of galaxies, we have compared our results with 
the predictions of recent models of chemical and dust evolution 
of galaxies with different star formation histories by 
\cite{Calura2008}. We conclude from this comparison that the
relations between $\mdust/\mstar$, $\mdust/\sfr$, \fmu\  and \ssfr\ 
could arise, at least in part, from an evolutionary sequence. As
galaxies form stars, their ISM becomes enriched in dust, while 
the drop in gas supply makes the specific star formation rate 
decrease. 
Interestingly, as a result, a young, actively star-forming
galaxy with low dust-to-gas ratio may still be highly dusty (in the
sense of a high $\mdust/\mstar$) because it contains large amounts
of interstellar gas (Fig.~\ref{fig:dust6}).
This may be important for the interpretation of the infrared emission
from young, gas-rich star-forming galaxies at high redshift.

The results presented in this paper should be especially useful
to improve the treatment of the ISM properties of galaxies in 
semi-analytic models of galaxy formation. More specifically, the
results of Section~\ref{dust:rel_sfr_dust} provide interesting 
constraints on the relation between star formation activity and 
dust content in such models. For example, the derived relation between 
\mdust\ and \sfr\ (Fig.~\ref{fig:dust5}) can be used in combination
of the Schmidt-Kennicutt law \citep{Schmidt1959, Kennicutt1998b} to 
describe the evolution of the dust content of galaxies together with
that of the stars and gas. Also, the correlation between \ssfr\ and
\fmu\ (Fig.~\ref{fig:dust6}) provides valuable clues on the way in 
which to implement the modelling of the different phases of the ISM
in simulated galaxies.

Our study provides a local reference for comparison with the properties
of galaxies at high redshifts. Observations with the {\it Spitzer} 
satellite have shown that infrared galaxies dominate the star formation
activity of the universe at redshift $z\sim1$ \citep{LeFloch2005}. 
Moreover, dust emission has been detected at sub-millimetre wavelengths
out to $z\sim6$, implying large dust masses and star formation rates of
the order of several $\times10^3\,\msun$\,yr$^{-1}$ in (at least some) 
young galaxies (e.g., \citealt{Chini1994,Isaak2002,Blain2002,Bertoldi2003,
Walter2009}). The production of such large amounts of dust on short 
timescales might require both efficient star formation and efficient grain 
condensation in supernova remnants (e.g., \citealt{Dunne2003,Morgan2003,
Maiolino2004}). This could affect the relation between star formation 
rate and dust mass. Forthcoming observations with the {\it Herschel
Space Telescope} and the {\it James Webb Space Telescope} will
enable the extension to high redshift of the type of analysis achieved
in this paper. This can only lead to an improved understanding of the
cosmic evolution of the dust content of galaxies. 

%******************************************************************************

\section*{acknowledgements}

We thank the referee, Charlie Conroy, for comments
and suggestions which helped improve the quality of this paper.
We are deeply grateful to David Schiminovich for matching the
SDSS-\iras\ sample with the latest \galex\ data release and for kindly
providing us with these data. We thank Jarle Brinchmann for useful
discussions and for providing us with the corrections to account 
for emission-line contamination in the optical SDSS bands. We also
thank Loretta Dunne, David Elbaz, Vivienne Wild, Jakob Walcher and Brent Groves
for useful discussions. We thank Francesco Calura and Antonio Pipino
for providing us with the electronic data of their chemical evolution
models. EdC was financed by
the EU Marie Curie Research Training Network MAGPOP and by the EU 
Marie Curie ToK Grant MTKD-CT-2006-039965.
CE was supported by the Marie Curie EARA-EST host fellowship
while this work was carried out and acknowledges the IAP for hospitality.
CE is partly supported by the Swiss Sunburst Fund.

Funding for the SDSS has been provided by the Alfred P. Sloan
Foundation, the Participating Institutions, the National Science
Foundation, the US Department of Energy, the National Aeronautics
and Space Administration, the Japanese Monbukagakusho, the Max
Planck Society, and the Higher Education Funding Council for
England. The SDSS Web site is http://www.sdss.org. The SDSS is
managed by the Astrophysical Research Consortium for the
Participating Institutions. The Participating Institutions are the
American Museum of Natural History, the Astrophysical Institute
Potsdam, the University of Basel, Cambridge University, Case
Western Reserve University, the University of Chicago, Drexel
University, Fermilab, the Institute for Advanced Study, the Japan
Participation Group, Johns Hopkins University, the Joint Institute
for Nuclear Astrophysics, the Kavli Institute for Particle
Astrophysics and Cosmology, the Korean Scientist Group, the
Chinese Academy of Sciences, Los Alamos National Laboratory, the
Max Planck Institute for Astronomy, the Max Planck Institute for
Astrophysics, New Mexico State University, Ohio State University,
the University of Pittsburgh, the University of Portsmouth,
Princeton University, the US Naval Observatory, and the University
of Washington.

%******************************************************************************

% Bibliography and bibfile
\def\aj{AJ}
\def\araa{ARA\&A}
\def\apj{ApJ}
\def\apjl{ApJ}
\def\apjs{ApJS}
\def\apss{Ap\&SS}
\def\aap{A\&A}
\def\aapr{A\&A~Rev.}
\def\aaps{A\&AS}
\def\mnras{MNRAS}
\def\pasp{PASP}
\def\pasj{PASJ}
\def\qjras{QJRAS}
\def\nat{Nature}

\def\aplett{Astrophys.~Lett.}
\def\aas{AAS}
\let\astap=\aap
\let\apjlett=\apjl
\let\apjsupp=\apjs
\let\applopt=\ao

\bibliographystyle{mn2e}
\bibliography{bib_dacunha}

\begin{thebibliography}{}

\bibitem[\protect\citeauthoryear{{Adelman-McCarthy et al.}}{{Adelman-McCarthy
  et al.}}{2008}]{Adelman_McCarthy2008}
{Adelman-McCarthy et al.} 2008, \apjs, 175, 297

\bibitem[\protect\citeauthoryear{{Armus et al.}}{{Armus et
  al.}}{2007}]{Armus2007}
{Armus et al.} 2007, \apj, 656, 148

\bibitem[\protect\citeauthoryear{{Baldwin}, {Phillips} \&
  {Terlevich}}{{Baldwin} et~al.}{1981}]{Baldwin1981}
{Baldwin} J.~A.,  {Phillips} M.~M.,    {Terlevich} R.,  1981, \pasp, 93, 5

\bibitem[\protect\citeauthoryear{{Beichman}, {Neugebauer}, {Habing}, {Clegg} \&
  {Chester}}{{Beichman} et~al.}{1988}]{Beichman1988}
{Beichman} C.~A.,  {Neugebauer} G.,  {Habing} H.~J.,  {Clegg} P.~E.,
  {Chester} T.~J.,  eds, 1988, {Infrared astronomical satellite (IRAS) catalogs
  and atlases. Volume 1: Explanatory supplement} Vol.~1

\bibitem[\protect\citeauthoryear{{Bertoldi}, {Carilli}, {Cox}, {Fan},
  {Strauss}, {Beelen}, {Omont} \& {Zylka}}{{Bertoldi}
  et~al.}{2003}]{Bertoldi2003}
{Bertoldi} F.,  {Carilli} C.~L.,  {Cox} P.,  {Fan} X.,  {Strauss} M.~A.,
  {Beelen} A.,  {Omont} A.,    {Zylka} R.,  2003, \aap, 406, L55

\bibitem[\protect\citeauthoryear{{Blain}, {Smail}, {Ivison}, {Kneib} \&
  {Frayer}}{{Blain} et~al.}{2002}]{Blain2002}
{Blain} A.~W.,  {Smail} I.,  {Ivison} R.~J.,  {Kneib} J.-P.,    {Frayer} D.~T.,
   2002, Physics Reports, 369, 111

\bibitem[\protect\citeauthoryear{{Blanton et al.}}{{Blanton et
  al.}}{2003}]{Blanton2003}
{Blanton et al.} 2003, \aj, 125, 2348

\bibitem[\protect\citeauthoryear{{Brandt}, {Hornschemeier}, {Alexander},
  {Garmire}, {Schneider}, {Broos}, {Townsley}, {Bautz}, {Feigelson} \&
  {Griffiths}}{{Brandt} et~al.}{2001}]{Brandt2001}
{Brandt} W.~N.,  {Hornschemeier} A.~E.,  {Alexander} D.~M.,  {Garmire} G.~P.,
  {Schneider} D.~P.,  {Broos} P.~S.,  {Townsley} L.~K.,  {Bautz} M.~W.,
  {Feigelson} E.~D.,    {Griffiths} R.~E.,  2001, \aj, 122, 1

\bibitem[\protect\citeauthoryear{{Brinchmann}, {Charlot}, {White}, {Tremonti},
  {Kauffmann}, {Heckman} \& {Brinkmann}}{{Brinchmann}
  et~al.}{2004}]{Brinchmann2004}
{Brinchmann} J.,  {Charlot} S.,  {White} S.~D.~M.,  {Tremonti} C.,  {Kauffmann}
  G.,  {Heckman} T.,    {Brinkmann} J.,  2004, \mnras, 351, 1151

\bibitem[\protect\citeauthoryear{{Bruzual}}{{Bruzual}}{2007}]{Bruzual2007}
{Bruzual} G.,  2007, (astro-ph/0703052)

\bibitem[\protect\citeauthoryear{{Bruzual} \& {Charlot}}{{Bruzual} \&
  {Charlot}}{2003}]{Bruzual2003}
{Bruzual} G.,  {Charlot} S.,  2003, \mnras, 344, 1000

\bibitem[\protect\citeauthoryear{{Calura}, {Pipino} \& {Matteucci}}{{Calura}
  et~al.}{2008}]{Calura2008}
{Calura} F.,  {Pipino} A.,    {Matteucci} F.,  2008, \aap, 479, 669

\bibitem[\protect\citeauthoryear{{Cao}, {Wu}, {Wang}, {Hao}, {Deng}, {Xia} \&
  {Zou}}{{Cao} et~al.}{2006}]{Cao2006}
{Cao} C.,  {Wu} H.,  {Wang} J.-L.,  {Hao} C.-N.,  {Deng} Z.-G.,  {Xia} X.-Y.,
   {Zou} Z.-L.,  2006, Chinese Journal of Astronomy and Astrophysics, 6, 197

\bibitem[\protect\citeauthoryear{{Chabrier}}{{Chabrier}}{2003}]{Chabrier2003}
{Chabrier} G.,  2003, \pasp, 115, 763

\bibitem[\protect\citeauthoryear{{Charlot} \& {Fall}}{{Charlot} \&
  {Fall}}{2000}]{Charlot2000}
{Charlot} S.,  {Fall} S.~M.,  2000, \apj, 539, 718

\bibitem[\protect\citeauthoryear{{Chiappini}, {Matteucci} \&
  {Gratton}}{{Chiappini} et~al.}{1997}]{Chiappini1997}
{Chiappini} C.,  {Matteucci} F.,    {Gratton} R.,  1997, \apj, 477, 765

\bibitem[\protect\citeauthoryear{{Chini} \& {Kruegel}}{{Chini} \&
  {Kruegel}}{1994}]{Chini1994}
{Chini} R.,  {Kruegel} E.,  1994, \aap, 288, L33

\bibitem[\protect\citeauthoryear{{Conroy}, {White} \& {Gunn}}{{Conroy}
  et~al.}{2009}]{Conroy2008}
{Conroy} C.,  {White} M.,    {Gunn} J.~E.,  2009, ArXiv e-prints

\bibitem[\protect\citeauthoryear{{da Cunha}, {Charlot} \& {Elbaz}}{{da Cunha}
  et~al.}{2008}]{daCunha2008}
{da Cunha} E.,  {Charlot} S.,    {Elbaz} D.,  2008, \mnras, 388, 1595

\bibitem[\protect\citeauthoryear{{de Grijp}, {Miley}, {Lub} \& {de Jong}}{{de
  Grijp} et~al.}{1985}]{DeGrijp1985}
{de Grijp} M.~H.~K.,  {Miley} G.~K.,  {Lub} J.,    {de Jong} T.,  1985, \nat,
  314, 240

\bibitem[\protect\citeauthoryear{{Dopita}, {Groves}, {Fischera}, {Sutherland},
  {Tuffs}, {Popescu}, {Kewley}, {Reuland} \& {Leitherer}}{{Dopita}
  et~al.}{2005}]{Dopita2005}
{Dopita} M.~A.,  {Groves} B.~A.,  {Fischera} J.,  {Sutherland} R.~S.,  {Tuffs}
  R.~J.,  {Popescu} C.~C.,  {Kewley} L.~J.,  {Reuland} M.,    {Leitherer} C.,
  2005, \apj, 619, 755

\bibitem[\protect\citeauthoryear{{Draine}, {Dale}, {Bendo} \& {et
  al.}}{{Draine} et~al.}{2007}]{Draine2007b}
{Draine} B.~T.,  {Dale} D.~A.,  {Bendo} G.,    {et al.} 2007, \apj, 663, 866

\bibitem[\protect\citeauthoryear{{Dudley} \& {Wynn-Williams}}{{Dudley} \&
  {Wynn-Williams}}{1997}]{Dudley1997}
{Dudley} C.~C.,  {Wynn-Williams} C.~G.,  1997, \apj, 488, 720

\bibitem[\protect\citeauthoryear{{Dunne}, {Eales}, {Edmunds}, {Ivison},
  {Alexander} \& {Clements}}{{Dunne} et~al.}{2000}]{Dunne2000}
{Dunne} L.,  {Eales} S.,  {Edmunds} M.,  {Ivison} R.,  {Alexander} P.,
  {Clements} D.~L.,  2000, \mnras, 315, 115

\bibitem[\protect\citeauthoryear{{Dunne}, {Eales}, {Ivison}, {Morgan} \&
  {Edmunds}}{{Dunne} et~al.}{2003}]{Dunne2003}
{Dunne} L.,  {Eales} S.,  {Ivison} R.,  {Morgan} H.,    {Edmunds} M.,  2003,
  \nat, 424, 285

\bibitem[\protect\citeauthoryear{{Dwek}}{{Dwek}}{1998}]{Dwek1998}
{Dwek} E.,  1998, \apj, 501, 643

\bibitem[\protect\citeauthoryear{{Dwek}, {Fioc} \& {Varosi}}{{Dwek}
  et~al.}{2000}]{Dwek2000}
{Dwek} E.,  {Fioc} M.,    {Varosi} F.,  2000, in {D.~Lemke, M.~Stickel, \&
  K.~Wilke} ed., ISO Survey of a Dusty Universe Vol.~548 of Lecture Notes in
  Physics, Berlin Springer Verlag, {The Effect of Dust Evolution on the
  Spectral Energy Distribution of Galaxies}.
pp 157--+

\bibitem[\protect\citeauthoryear{{Edmunds}}{{Edmunds}}{2001}]{Edmunds2001}
{Edmunds} M.~G.,  2001, \mnras, 328, 223

\bibitem[\protect\citeauthoryear{{Elbaz}, {Cesarsky}, {Chanial}, {Aussel},
  {Franceschini}, {Fadda} \& {Chary}}{{Elbaz} et~al.}{2002}]{Elbaz2002}
{Elbaz} D.,  {Cesarsky} C.~J.,  {Chanial} P.,  {Aussel} H.,  {Franceschini} A.,
   {Fadda} D.,    {Chary} R.~R.,  2002, \aap, 384, 848

\bibitem[\protect\citeauthoryear{{Helou}, {Khan}, {Malek} \& {Boehmer}}{{Helou}
  et~al.}{1988}]{Helou1988}
{Helou} G.,  {Khan} I.~R.,  {Malek} L.,    {Boehmer} L.,  1988, \apjs, 68, 151

\bibitem[\protect\citeauthoryear{{Hildebrand}}{{Hildebrand}}{1983}]{Hildebrand%
1983}
{Hildebrand} R.~H.,  1983, \qjras, 24, 267

\bibitem[\protect\citeauthoryear{{Hopkins}, {Connolly}, {Haarsma} \&
  {Cram}}{{Hopkins} et~al.}{2001}]{Hopkins2001}
{Hopkins} A.~M.,  {Connolly} A.~J.,  {Haarsma} D.~B.,    {Cram} L.~E.,  2001,
  \aj, 122, 288

\bibitem[\protect\citeauthoryear{{Hornschemeier}, {Brandt}, {Garmire},
  {Schneider}, {Barger}, {Broos}, {Cowie}, {Townsley}, {Bautz}, {Burrows},
  {Chartas}, {Feigelson}, {Griffiths}, {Lumb}, {Nousek}, {Ramsey} \&
  {Sargent}}{{Hornschemeier} et~al.}{2001}]{Hornschemeier2001}
{Hornschemeier} A.~E.,  {Brandt} W.~N.,  {Garmire} G.~P.,  {Schneider} D.~P.,
  {Barger} A.~J.,  {Broos} P.~S.,  {Cowie} L.~L.,  {Townsley} L.~K.,  {Bautz}
  M.~W.,  {Burrows} D.~N.,  {Chartas} G.,  {Feigelson} E.~D.,  {Griffiths}
  R.~E.,  {Lumb} D.,  {Nousek} J.~A.,  {Ramsey} L.~W.,    {Sargent} W.~L.~W.,
  2001, \apj, 554, 742

\bibitem[\protect\citeauthoryear{{Imanishi} \& {Dudley}}{{Imanishi} \&
  {Dudley}}{2000}]{Imanishi2000}
{Imanishi} M.,  {Dudley} C.~C.,  2000, \apj, 545, 701

\bibitem[\protect\citeauthoryear{{Isaak}, {Priddey}, {McMahon}, {Omont},
  {Peroux}, {Sharp} \& {Withington}}{{Isaak} et~al.}{2002}]{Isaak2002}
{Isaak} K.~G.,  {Priddey} R.~S.,  {McMahon} R.~G.,  {Omont} A.,  {Peroux} C.,
  {Sharp} R.~G.,    {Withington} S.,  2002, \mnras, 329, 149

\bibitem[\protect\citeauthoryear{{Isobe}, {Feigelson}, {Akritas} \&
  {Babu}}{{Isobe} et~al.}{1990}]{Isobe1990}
{Isobe} T.,  {Feigelson} E.~D.,  {Akritas} M.~G.,    {Babu} G.~J.,  1990, \apj,
  364, 104

\bibitem[\protect\citeauthoryear{{Issa}, {MacLaren} \& {Wolfendale}}{{Issa}
  et~al.}{1990}]{Issa1990}
{Issa} M.~R.,  {MacLaren} I.,    {Wolfendale} A.~W.,  1990, \aap, 236, 237

\bibitem[\protect\citeauthoryear{{Kauffmann et al.}}{{Kauffmann et
  al.}}{2003a}]{Kauffmann2003a}
{Kauffmann et al.} 2003a, \mnras, 341, 33

\bibitem[\protect\citeauthoryear{{Kauffmann et al.}}{{Kauffmann et
  al.}}{2003b}]{Kauffmann2003b}
{Kauffmann et al.} 2003b, \mnras, 341, 54

\bibitem[\protect\citeauthoryear{{Kennicutt}
  Jr.}{{Kennicutt}}{1998a}]{Kennicutt1998b}
{Kennicutt} Jr. R.~C.,  1998a, \araa, 36, 189

\bibitem[\protect\citeauthoryear{{Kennicutt}
  Jr.}{{Kennicutt}}{1998b}]{Kennicutt1998}
{Kennicutt} Jr. R.~C.,  1998b, \apj, 498, 541

\bibitem[\protect\citeauthoryear{{Kennicutt} Jr., {Armus}, {Bendo}, {Calzetti},
  {Dale}, {Draine}, {Engelbracht}, {Gordon} \& {et al.}}{{Kennicutt}
  et~al.}{2003}]{Kennicutt2003}
{Kennicutt} Jr. R.~C.,  {Armus} L.,  {Bendo} G.,  {Calzetti} D.,  {Dale} D.~A.,
   {Draine} B.~T.,  {Engelbracht} C.~W.,  {Gordon} K.~D.,    {et al.} 2003,
  \pasp, 115, 928

\bibitem[\protect\citeauthoryear{{Kewley}, {Jansen} \& {Geller}}{{Kewley}
  et~al.}{2005}]{Kewley2005}
{Kewley} L.~J.,  {Jansen} R.~A.,    {Geller} M.~J.,  2005, \pasp, 117, 227

\bibitem[\protect\citeauthoryear{{Kong}, {Charlot}, {Brinchmann} \&
  {Fall}}{{Kong} et~al.}{2004}]{Kong2004}
{Kong} X.,  {Charlot} S.,  {Brinchmann} J.,    {Fall} S.~M.,  2004, \mnras,
  349, 769

\bibitem[\protect\citeauthoryear{{Le Floc'h}, {Papovich}, {Dole} \& {et
  al.}}{{Le Floc'h} et~al.}{2005}]{LeFloch2005}
{Le Floc'h} E.,  {Papovich} C.,  {Dole} H.,    {et al.} 2005, \apj, 632, 169

\bibitem[\protect\citeauthoryear{{Lisenfeld} \& {Ferrara}}{{Lisenfeld} \&
  {Ferrara}}{1998}]{Lisenfeld1998}
{Lisenfeld} U.,  {Ferrara} A.,  1998, \apj, 496, 145

\bibitem[\protect\citeauthoryear{{Maiolino}, {Schneider}, {Oliva}, {Bianchi},
  {Ferrara}, {Mannucci}, {Pedani} \& {Roca Sogorb}}{{Maiolino}
  et~al.}{2004}]{Maiolino2004}
{Maiolino} R.,  {Schneider} R.,  {Oliva} E.,  {Bianchi} S.,  {Ferrara} A.,
  {Mannucci} F.,  {Pedani} M.,    {Roca Sogorb} M.,  2004, \nat, 431, 533

\bibitem[\protect\citeauthoryear{{Maller}, {Berlind}, {Blanton} \&
  {Hogg}}{{Maller} et~al.}{2009}]{Maller2009}
{Maller} A.~H.,  {Berlind} A.~A.,  {Blanton} M.~R.,    {Hogg} D.~W.,  2009,
  \apj, 691, 394

\bibitem[\protect\citeauthoryear{{Marigo} \& {Girardi}}{{Marigo} \&
  {Girardi}}{2007}]{Marigo2007}
{Marigo} P.,  {Girardi} L.,  2007, \aap, 469, 239

\bibitem[\protect\citeauthoryear{{Martin et al.}}{{Martin et
  al.}}{2005}]{Martin2005}
{Martin et al.} 2005, \apjl, 619, L1

\bibitem[\protect\citeauthoryear{{Morgan} \& {Edmunds}}{{Morgan} \&
  {Edmunds}}{2003}]{Morgan2003}
{Morgan} H.~L.,  {Edmunds} M.~G.,  2003, \mnras, 343, 427

\bibitem[\protect\citeauthoryear{{Morrissey et al.}}{{Morrissey et
  al.}}{2005}]{Morrissey2005}
{Morrissey et al.} 2005, \apjl, 619, L7

\bibitem[\protect\citeauthoryear{{Moshir}}{{Moshir}}{1989}]{Moshir1989}
{Moshir} M.,  1989, {IRAS Faint Source Survey, Explanatory supplement version 1
  and tape}.
Pasadena: Infrared Processing and Analysis Center, California Institute of
  Technology, 1989, edited by Moshir, M.

\bibitem[\protect\citeauthoryear{{Obri{\'c}}, {Ivezi{\'c}}, {Best}, {Lupton},
  {Tremonti}, {Brinchmann}, {Ag{\"u}eros} \& {et al.}}{{Obri{\'c}}
  et~al.}{2006}]{Obric2006}
{Obri{\'c}} M.,  {Ivezi{\'c}} {\v Z}.,  {Best} P.~N.,  {Lupton} R.~H.,
  {Tremonti} C.,  {Brinchmann} J.,  {Ag{\"u}eros} M.~A.,    {et al.} 2006,
  \mnras, 370, 1677

\bibitem[\protect\citeauthoryear{{Pasquali}, {Kauffmann} \&
  {Heckman}}{{Pasquali} et~al.}{2005}]{Pasquali2005}
{Pasquali} A.,  {Kauffmann} G.,    {Heckman} T.~M.,  2005, \mnras, 361, 1121

\bibitem[\protect\citeauthoryear{{Pipino}, {Kawata}, {Gibson} \&
  {Matteucci}}{{Pipino} et~al.}{2005}]{Pipino2005}
{Pipino} A.,  {Kawata} D.,  {Gibson} B.~K.,    {Matteucci} F.,  2005, \aap,
  434, 553

\bibitem[\protect\citeauthoryear{{Rigopoulou et al.}}{{Rigopoulou et
  al.}}{1999}]{Rigopoulou1999}
{Rigopoulou et al.} 1999, in {Cox} P.,  {Kessler} M.,  eds, The Universe as
  Seen by ISO Vol.~427 of ESA Special Publication, {Ultraluminous IRAS galaxies
  as seen with ISO}.
pp 833--+

\bibitem[\protect\citeauthoryear{{Salim}, {Dickinson}, {Rich}, {Charlot},
  {Lee}, {Schiminovich}, {Perez-Gonzalez}, {Ashby}, {Papovich}, {Faber},
  {Ivison}, {Frayer}, {Walton}, {Weiner}, {Chary}, {Bundy}, {Noeske} \&
  {Koekemoer}}{{Salim} et~al.}{2009}]{Salim2009}
{Salim} S.,  {Dickinson} M.,  {Rich} R.~M.,  {Charlot} S.,  {Lee} J.~C.,
  {Schiminovich} D.,  {Perez-Gonzalez} P.~G.,  {Ashby} M.~L.~N.,  {Papovich}
  C.,  {Faber} S.~M.,  {Ivison} R.~J.,  {Frayer} D.~T.,  {Walton} J.~M.,
  {Weiner} B.~J.,  {Chary} R.-R.,  {Bundy} K.,  {Noeske} K.,    {Koekemoer}
  A.~M.,  2009, ArXiv e-prints

\bibitem[\protect\citeauthoryear{{Salpeter}}{{Salpeter}}{1955}]{Salpeter1955}
{Salpeter} E.~E.,  1955, \apj, 121, 161

\bibitem[\protect\citeauthoryear{{Saunders et al.}}{{Saunders et
  al.}}{2000}]{Saunders2000}
{Saunders et al.} 2000, \mnras, 317, 55

\bibitem[\protect\citeauthoryear{{Schlegel}, {Finkbeiner} \&
  {Davis}}{{Schlegel} et~al.}{1998}]{Schlegel1998}
{Schlegel} D.~J.,  {Finkbeiner} D.~P.,    {Davis} M.,  1998, \apj, 500, 525

\bibitem[\protect\citeauthoryear{{Schmidt} \& {Boller}}{{Schmidt} \&
  {Boller}}{1993}]{Schmidt1993}
{Schmidt} K.-H.,  {Boller} T.,  1993, Astronomische Nachrichten, 314, 361

\bibitem[\protect\citeauthoryear{{Schmidt}}{{Schmidt}}{1959}]{Schmidt1959}
{Schmidt} M.,  1959, \apj, 129, 243

\bibitem[\protect\citeauthoryear{{Seibert et al.}}{{Seibert et
  al.}}{2005}]{Seibert2005}
{Seibert et al.} 2005, \apjl, 619, L55

\bibitem[\protect\citeauthoryear{{Shimasaku}, {Fukugita}, {Doi}, {Hamabe},
  {Ichikawa}, {Okamura}, {Sekiguchi}, {Yasuda}, {Brinkmann}, {Csabai},
  {Ichikawa}, {Ivezi{\'c}}, {Kunszt}, {Schneider}, {Szokoly}, {Watanabe} \&
  {York}}{{Shimasaku} et~al.}{2001}]{Shimasaku2001}
{Shimasaku} K.,  {Fukugita} M.,  {Doi} M.,  {Hamabe} M.,  {Ichikawa} T.,
  {Okamura} S.,  {Sekiguchi} M.,  {Yasuda} N.,  {Brinkmann} J.,  {Csabai} I.,
  {Ichikawa} S.,  {Ivezi{\'c}} Z.,  {Kunszt} P.~Z.,  {Schneider} D.~P.,
  {Szokoly} G.~P.,  {Watanabe} M.,    {York} D.~G.,  2001, \aj, 122, 1238

\bibitem[\protect\citeauthoryear{{Silva}, {Granato}, {Bressan} \&
  {Danese}}{{Silva} et~al.}{1998}]{Silva1998}
{Silva} L.,  {Granato} G.~L.,  {Bressan} A.,    {Danese} L.,  1998, \apj, 509,
  103

\bibitem[\protect\citeauthoryear{{Skrutskie et al.}}{{Skrutskie et
  al.}}{2006}]{Skrutskie2006}
{Skrutskie et al.} 2006, \aj, 131, 1163

\bibitem[\protect\citeauthoryear{{Soifer}, {Sanders}, {Madore}, {Neugebauer},
  {Danielson}, {Elias}, {Lonsdale} \& {Rice}}{{Soifer}
  et~al.}{1987}]{Soifer1987}
{Soifer} B.~T.,  {Sanders} D.~B.,  {Madore} B.~F.,  {Neugebauer} G.,
  {Danielson} G.~E.,  {Elias} J.~H.,  {Lonsdale} C.~J.,    {Rice} W.~L.,  1987,
  \apj, 320, 238

\bibitem[\protect\citeauthoryear{{Strateva et al.}}{{Strateva et
  al.}}{2001}]{Strateva2001}
{Strateva et al.} 2001, \aj, 122, 1861

\bibitem[\protect\citeauthoryear{{Sullivan}, {Mobasher}, {Chan}, {Cram},
  {Ellis}, {Treyer} \& {Hopkins}}{{Sullivan} et~al.}{2001}]{Sullivan2001}
{Sullivan} M.,  {Mobasher} B.,  {Chan} B.,  {Cram} L.,  {Ellis} R.,  {Treyer}
  M.,    {Hopkins} A.,  2001, \apj, 558, 72

\bibitem[\protect\citeauthoryear{{Tremonti et al.}}{{Tremonti et
  al.}}{2004}]{Tremonti2004}
{Tremonti et al.} 2004, \apj, 613, 898

\bibitem[\protect\citeauthoryear{{Veilleux}, {Kim} \& {Sanders}}{{Veilleux}
  et~al.}{1999}]{Veilleux1999}
{Veilleux} S.,  {Kim} D.-C.,    {Sanders} D.~B.,  1999, \apj, 522, 113

\bibitem[\protect\citeauthoryear{{Veilleux}, {Kim}, {Sanders}, {Mazzarella} \&
  {Soifer}}{{Veilleux} et~al.}{1995}]{Veilleux1995}
{Veilleux} S.,  {Kim} D.-C.,  {Sanders} D.~B.,  {Mazzarella} J.~M.,    {Soifer}
  B.~T.,  1995, \apjs, 98, 171

\bibitem[\protect\citeauthoryear{{Walter}}{{Walter}}{2009}]{Walter2009}
{Walter} F.,  2009, ArXiv e-prints

\bibitem[\protect\citeauthoryear{{Wang} \& {Heckman}}{{Wang} \&
  {Heckman}}{1996}]{Wang1996}
{Wang} B.,  {Heckman} T.~M.,  1996, \apj, 457, 645

\bibitem[\protect\citeauthoryear{{Werner}, {Roellig}, {Low}, {Rieke}, {Rieke},
  {Hoffmann}, {Young} \& {Houck}}{{Werner} et~al.}{2004}]{Werner2004}
{Werner} M.~W.,  {Roellig} T.~L.,  {Low} F.~J.,  {Rieke} G.~H.,  {Rieke} M.,
  {Hoffmann} W.~F.,  {Young} E.,    {Houck} J.~R.,  2004, \apjs, 154, 1

\end{thebibliography}

\end{document}